\documentclass[
 aip,
 floatfix,
 amsmath,amssymb,
 reprint,
]{revtex4-1}

\usepackage{graphicx}
\usepackage{dcolumn}
\usepackage{bm}

\usepackage{subcaption}
\usepackage[utf8]{inputenc}
\usepackage[T1]{fontenc}
\usepackage{mathptmx}
\usepackage{etoolbox}

\begin{document}

\title[Elucidating coherent structures, transport barriers and  entrainment in turbulent fountains in stratified media]{Elucidating coherent structures, transport barriers and  entrainment in turbulent fountains in stratified media}

\author{D. Freire Caporale}
    \email{dfreire@fisica.edu.uy}
    \affiliation{Instituto de Física, Facultad de Ciencias, Universidad de la República, Uruguay}

\author{N. Barrere}%
    \affiliation{Departamento de Física, Centro Universitario Regional Este,  Universidad de la República, Uruguay}

\author{Arturo C. Martí}
    \affiliation{Instituto de Física, Facultad de Ciencias, Universidad de la República, Uruguay}
    
\author{C. Cabeza}
    \affiliation{Instituto de Física, Facultad de Ciencias, Universidad de la República, Uruguay}
    
\author{L. G. Sarasúa}
    \affiliation{Instituto de Física, Facultad de Ciencias, Universidad de la República, Uruguay}

\date{\today}

\begin{abstract}
We analyse the flow organization of turbulent fountains in stratified media under different conditions, using three-dimensional finite-time Lyapunov exponents. The dominant Lagrangian coherent structures responsible for the transport barriers in three different configurations suggest a self-similarity behaviour. After proposing a criterion for delimiting the boundary surface of the uprising fountain, we quantify the entrainment and re-entrainment rates under fully developed flow conditions using the proper coefficients. Finally, our analysis was applied to the Selective Inverted Sink, a technological application of turbulent fountains, identifying turbulence as the primary mechanism favouring the device's efficiency.
\end{abstract}

\maketitle

\section{\label{sec:intro} Introduction}
A fountain is a vertical buoyant jet in which the buoyancy force and the jet’s initial velocity act in opposite directions. The flow is a plume if the buoyancy force acts in the same direction as the jet velocity. Fountains and plumes are encountered frequently in nature as well as in technical applications. Since the dynamic behaviour of fluids within stratified media presents a problem of considerable interest across a number of fields, turbulent fountains and plumes in both uniform and stratified mediums have been the subject of research for decades~\cite{turner1969buoyant, baines1990turbulent, bloomfield1998turbulent, bloomfield1999turbulent, lin2000direct, hunt2001virtual, hunt2005lazy, kaye2008turbulent, woods2010turbulent, richards2014radial, hunt2015fountains, carroll2015modeling, camassa2016optimal, ezhova2016interaction}. The fountain dynamics in a stratified medium can be outlined as follows. At an initial stage, the fountain decelerates due to the entrainment of ambient fluid and the opposing buoyancy force, reaching a maximum height in which the momentum is zero. Then the flow reverses direction and falls as an annular plume around the fountain core. Depending on the initial fluxes of momentum and buoyancy and the stratification profile, the fountain spreads outwards at a non-zero spreading height above the source level, or the flow collapses, i.e., it falls to the source level.

The theoretical description of turbulent fountains in the quasi-steady regime can be done based on the influential work of Morton, Taylor and Turner~\cite{morton1956turbulent, morton1959forced, morton1973scale}, who derived the so-called MTT equations for the evolution of volume, momentum, and buoyancy fluxes in fountains. In deriving these equations, it is assumed that the horizontal velocity at which the ambient fluid enters the fountain is proportional to the vertical velocity in the fountain, with a proportionality constant $\alpha$ called entrainment coefficient. Although successful in predicting the evolution in a uniform ambient or the maximum height in plumes~\cite{kaye2008turbulent}, the MTT equations do not describe the dynamics after the vertical velocity reverses its direction.

Bloomfield and Kerr proposed that the spreading height, $z_s$, can be obtained matching it to the height where the fluid of the environment has the density of the fluid at the maximum height, $z_m$, and they used this condition to obtain estimations of $z_s$ and $z_m$ combining different models~\cite{bloomfield1998turbulent}. This may be considered a first order estimation because the mentioned condition does not take into account the mixing between the jet and ambient fluids in the downflow that occurs after the fountain reverses direction. In a later work, Bloomfield and Kerr developed a theoretical model to predict the maximum spreading height~\cite{bloomfield2000theoretical} based on the equations derived by McDougall for an axisymmetric fountain in a homogeneous fluid~\cite{mcdougall1981negatively}. In this model, the authors assumed entrainment equations that depend on the model parameters.

Some years later, Kaminski et al.~\cite{kaminski2005turbulent} developed an expression for the entrainment parameter depending on three parameters that can be determined using experimental data. A comparison between the predictions based on this expression and the experimental data were given in Ref.~\onlinecite{carazzo2010rise} for the case of homogeneous mediums. Mehaddi et al.~\cite{mehaddi2012analytical} conducted a study of fountains in stratified environments and obtained expression for the maximum height, although the spreading behaviour of the fountain was not considered in this investigation. Papanicolau et al.~\cite{papanicolaou2010spreading} conducted an experimental study on the collapse and spreading of turbulent fountains and performed a comparison with those obtained in Ref.~\onlinecite{bloomfield1998turbulent}. As various authors have pointed out~\cite{turner_1973, telford1966convective, van2015energy}, to assume a constant $\alpha$ is an approximation because the entrainment coefficient depends on the turbulence intensity, and, as a consequence, it can vary with the rise of the fountain. Recently, Sarasua et al.~\cite{sarasua2021spreading} proposed a model that generalizes those of Morton et al. in order to determine the dependence of the maximum height and the spreading height with the parameters involved. This model determines the critical conditions for the collapse of the fountain, i.e., when the jet falls to the source level, using a parameter that measures the mixing of the jet with the environment along the downflow. The value of this parameter has been estimated using numerical simulations.

The present work is focused on analysing in detail the flow structure in the downflow to improve the knowledge of the mechanisms involved in the source dynamics from a Lagrangian perspective. For this purpose, we performed numerical simulations and studied the Lagrangian Coherent Structures (LCSs) of the fully developed flow using Finite-Time Lyapunov exponents (FTLE). In the light of this analysis, we propose a criterion for quantifying the entrainment and re-entrainment. The organisation of the rest of this work is as follows. In Sec.~\ref{sec:problem}, we describe the problem of turbulent fountains in stratified media and provide a summary of its state-of-the-art. In Sec.~\ref{sec:ftle}, we mention the main concepts related to the FTLE analysis, which is the basis of our study. Sec.~\ref{sec:measurement} is devoted to describing the numerical simulations and the laboratory experiments that we performed, as well as their validation. In Sec.~\ref{sec:results}, we present the results and an in-depth analysis, which includes the discovery of LCSs that organise the flow, a criterion for quantifying the mixing, and its application to the study of the efficiency of an innovative technological device such as the Selective Inverted Sink (SIS)~\cite{guarga2000evaluation}. Finally, we summarize the conclusions in Sec.~\ref{sec:conclusions}.

\section{\label{sec:problem} Problem description}
Fountains and plumes play an essential role in the dispersion of pollutants released into the air by natural phenomena or industrial applications. The sources that discharge material into the environment include volcanic eruptions, sewage ocean outfalls and chimney factories. Consequently, it is of interest to develop models that are able to predict the final altitude of the plume. Bloomfield and Kerr proposed that the spreading height, $z_s$, can be obtained matching it to the height where the fluid of the environment has the density of the fluid at the maximum height~\cite{bloomfield1998turbulent}.

This estimate is very crude because the mentioned condition does not consider mixing the jet and ambient fluids in the downflow after the fountain reverses direction. Sarasua et al.~\cite{sarasua2021spreading} introduced a parameter $\gamma$ that represents the proportion of environment fluid to jet fluid that mixes along the downflow to form the fountain fluid at the spreading region. The value of this quantity has been estimated from numerical simulations. Ref.~\onlinecite{sarasua2021spreading} shows that using this parameter and a set of equations that generalizes those of Morton et al., we obtain values of $z_s$ and $z_m$ that are in good agreement with the experiments. The quantity $\gamma$ is an effective parameter that reflects the total effect of the mixing, and so it serves to determine the final quasi-stationary stage. Therefore, it is desirable to obtain equations to describe the details of the flow in time and space.

Bloomfield and Kerr~\cite{bloomfield2000theoretical} developed a theoretical model to predict the maximum spreading height based on the equations derived by McDougall~\cite{mcdougall1981negatively} for an axisymmetric fountain in a homogeneous fluid. In this model, the authors assumed entrainment equations that depend on three entrainment constants and represent mixing between different parts of the fountain. The drawbacks of these models reside in the many assumptions that they make. In particular, they introduce several entrainment parameters whose values are known. In addition, the validity of the proper entrainment hypothesis between different parts of the fountain is not apparent. The present work aims to study the flow structure in detail to test different hypotheses and estimate parameters values characterizing the entrainment.

\section{\label{sec:ftle} Finite-Time Lyapunov Exponents}
The Finite-Time Lyapunov exponents (FTLE) is a scalar field that measures the exponential growth of the distance between initially close particles during the time interval $(t_0,t_0+\tau)$~\cite{haller_2002,haller2015}. That is, if $\left \| \delta \mathbf{x}_0 \right \|$ is the initial separation of two fluid particles, the maximum separation is given by
\begin{equation}
\left \| \delta \mathbf{x} \right \|_{\textrm{max}}=e^{\left | \tau \right |\Lambda_{t_0,\tau}}\left \| \delta \mathbf{x}_0 \right \|
\label{ftle}
\end{equation}
where $\Lambda_{t_0,\tau}$ is the FTLE field calculated in the time interval $(t_0, t_0+\tau)$. The works of Haller~\cite{haller_2000,haller_2000_chaos} define Lagrangian Coherent Structures as manifolds that acts as separatrices of the flow, i.e., material barriers of the flow. Shadden~\cite{shadden_2005,shadden-exp} demonstrated that the ridges of the FTLE field correspond to LCS. Then, ridges of FTLE behave as material barriers that govern the flow dynamics.

For the FTLE calculation, particle trajectories $\mathbf{x}(t)$ for times $t\in(t_0,t_0+\tau)$ need to be estimated. This is done based on the velocity data. Hence, from the Eulerian 3-dimensional velocity fields $\mathbf{v}(\mathbf{x}(t),t)$, we solved the equation numerically
\begin{equation}
\mathbf{\dot{x}}\left ( t \right )=\mathbf{v}\left( \mathbf{x}, t \right )
\label{eq:flux}
\end{equation} 
To recover the dominant LCS that organises the flow, first we smoothed (or filtered) its inherent turbulent fluctuations by taking the azimuthal average of the computed 3-dimensional velocity fields, interpolated into many different vertical planes (about 50 planes), containing the fountain axis (i.e., the $z$-axis) and uniformly distributed around the axis. In fact, at stages where the flow is fully developed, the flow is observed to be roughly stationary away from the jet axis (since the fluctuations are weak).

Moreover, by filtering the 3-dimensional tracer field (e.g., Fig.~\ref{fig:tinta_com}) in the same way and setting a suitable threshold level for the tracer field, we defined a criterion to draw the \textit{contour} of the flow. In Fig.~\ref{fig:trayectorias_sobre_FTLE_G15} such contour is shown with a dashed magenta line.

Once the 3-dimensional velocity field is smoothed into a 2-dimensional velocity field (i.e., in the plane $(x,z)$), we obtained a 2-dimensional mesh (cartesian in our case) where the velocities were computed at every time step. Within each calculation cell, at time $t=0$, we located three fluid particles along each direction of the cell (i.e., $3 \times 3$ fluid particles within each numerical cell). Then, those particles were advected, following a fourth order Runge-Kutta scheme, and cubic interpolation was used to compute particle velocity through the domain. Once the particle trajectories are obtained, FTLE fields are computed using the Cauchy-Green tensor $C$. The quantity $\left \| \delta \mathbf{x} \right \|_{\textrm{max}}$ is aligned with the eigenvector associated with the maximum eigenvalue, $\lambda_{\textrm{max}}(C)$ of tensor $C$, leading to $\left \| \delta \mathbf{x} \right \|_{\textrm{max}}=\sqrt{\lambda_{\textrm{max}}(C)}\left \| \delta \mathbf{x}_0 \right \|$. Then, the FTLE field is computed from the following:
\begin{equation}
\Lambda_{t_0,\tau}=\frac{1}{\left | \tau \right |}ln(\sqrt{\lambda_{\textrm{max}}(C)}),
\label{ftle_practica}
\end{equation}
where the absolute value of $\tau$ is taken since the particle trajectories can be advected forward in time ($\tau>0$) and backward in time ($\tau <0$). The FTLE field of particles forward in time is called $\Lambda_{t_0,\tau}^+$ and reveals repelling manifolds (repelling LCS). On the other hand, the FTLE field of particles advected backwards in time is called $\Lambda_{t_0,\tau}^-$ and reveals attracting manifolds. In this work, we only show $\Lambda_{t_0,\tau}^+$ in order to extract repelling manifolds (attracting LCS).

The choice of the length of time over which the FTLE is computed
is critical to reveal different LCSs~\cite{haller2015}. We chose $\tau=25$~s since it is a timelapse, which spans the most important processes during the development of the flow. For instance, for all the configurations considered, the flow reaches its maximum height within the first 25 seconds, and the spreading flow is already formed. Moreover, our study considered several values and observed that FTLE fields remain unchanged, taking $\tau=25$~s or longer.

\section{\label{sec:measurement} Numerical simulations}
In our work, we performed three-dimensional numerical simulations of turbulent fountains under different configurations. We analysed the scalar fields of temperature and the vectorial velocity field obtained, previously validated by comparison with laboratory experiments from Freire et al.~\cite{freire2010effect}, and we tracked the inflow using an ink tracer. The following sections~\ref{subsec:simulations} and \ref{subsubsec:validation} explain the numerical calculation software and its validation with experimental measurements, respectively.

\subsection{Experimental setup}\label{subsec:experiments}

We performed the laboratory experiments using the experimental setup described in Freire et al.~\cite{freire2015formation} (see also Ref.~\onlinecite{freire2010effect}) which consists of a water jet that enters  vertically from below into a prismatic container through a circular nozzle. The container is initially filled with water (ambient fluid) linearly stratified in temperature and height. The inlet port diameter is $D=8$~mm, and the container dimensions are 0.4~m on each lateral side ($-0.2$~m~$\le x, \ y \le$~$+0.2$~m) and 1~m in height ($0\le z \le 1$~m). The ambient fluid temperature was set at 15~$^\circ$C at the bottom ($z=0$) and 40~$^\circ$C at the top ($z=1$~m). The jet temperature ($T_{jet}$) and we set the flow rate at the inlet at 15~$^\circ$C and 5.5~cm$^3$s$^{-1}$, respectively.

Experimentally, we studied an additional configuration using a stainless-steel wire mesh placed at the inlet port, which increases the turbulence of the fountain at the entrance. We named the configuration with and without said mesh as G-15 and F-15, respectively.

\begin{figure}[hbt]
\hspace{1pc}
\centering
\begin{subfigure}{0.4\linewidth}
    \includegraphics[height=6pc]{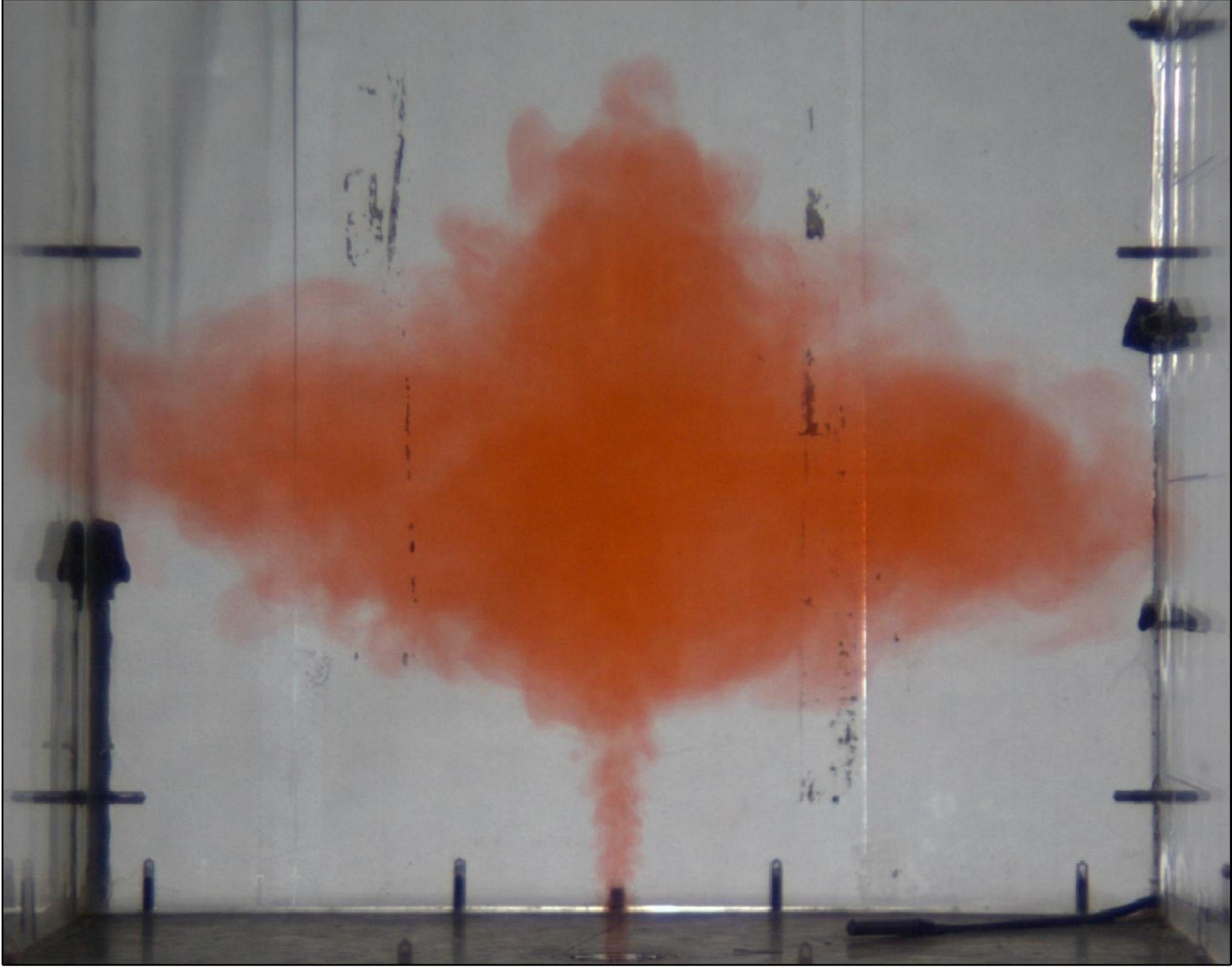}
    \caption{Experiment}
    \label{fig:tinta_exp}
\end{subfigure}
\hspace{0.05pc} 
\begin{subfigure}{0.5\linewidth}
    \includegraphics[height=6pc]{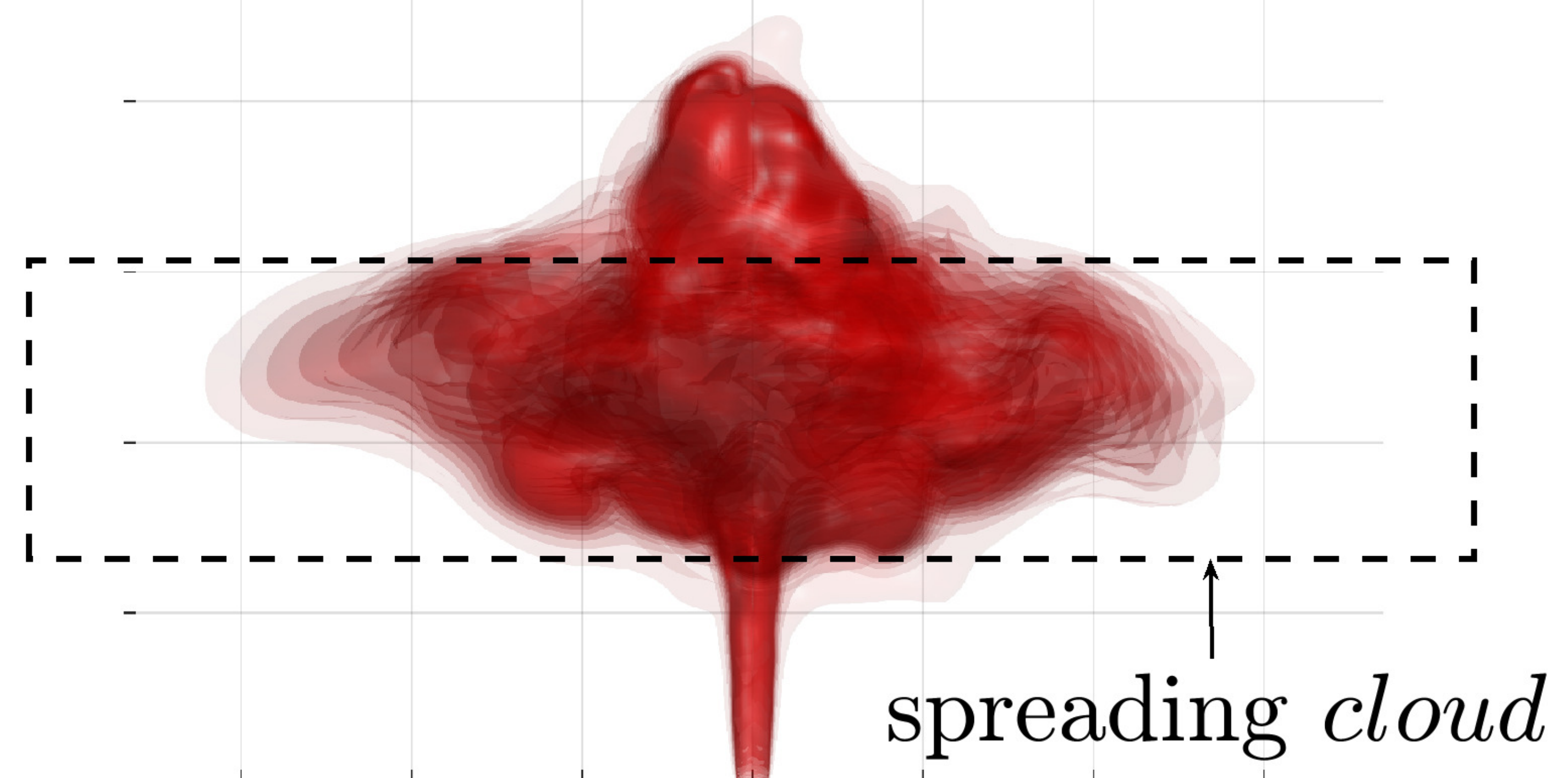}
    \caption{Simulation}
    \label{fig:tinta_com}
\end{subfigure}
\caption{Visualisation of the flow with an ink tracer for the F-15 configuration at $t=80$~s. In (b), we indicate the region of the flow herein referred to as the spreading \textit{cloud}.}
\label{fig:validation_ink}
\end{figure}

An ink tracer was added to the inlet jet to visualize the developed flow, to obtain images like the one shown in Fig.~\ref{fig:tinta_exp}. A Digital Particle Image Velocimetry (DPIV)~\cite{adrian1984scattering} was performed to measure 2D velocity fields at a vertical plane along the fountain axis, as shown in Figs.~\ref{fig:ux_exp} and \ref{fig:uz_exp}. It should be noted that, due to the limitations of the CMOS digital camera, we could not measure the velocity along the fountain axis, but rather outside it.

\subsection{Numerical scheme}\label{subsec:simulations}

We performed the numerical simulations with the open-source package caffa3d.MBRi~\cite{usera2008parallel,mendina2014general}, which implements a fully implicit 3D incompressible Finite Volume Method (FVM) with second-order accuracy in space and time and uses curvilinear meshes aimed at Navier–Stokes equations in complex geometries. It has been validated and tested for accuracy, exhibiting mesh and time independence in benchmark flows. Turbulence implementation follows the standard Smagorinsky large-eddy simulation model~\cite{smagorinsky1963general}.

In our simulations, the time step was 0.05~s, and the total number of mesh nodes was $8\times 10^6$. Mesh independence was verified using a coarser and a finer grid of $5\times10^6$ and $30\times10^6$ nodes.

We modelled the ink tracer from the experiments as a passive scalar field and the turbulence generation due to the wire mesh as white noise in the jet velocity at the inlet. In the absence of such disturbance, the velocity at the inlet was set vertically with a top-hat
profile. The intensity of the noise was varied to obtain the best fit of experimental observations. Noise levels of 0.2\% and 20\% of the velocity magnitude at the inlet are in full agreement with experiments F-15 and G-15, respectively, as we describe in the next section. Additionally, the F-18 configuration, which is equal to the F-15 configuration but with $T_{jet}=18~^\circ$C, was considered in the simulations.

The following section compares the experimental results, for the G-15 and F-15 configurations, with simulations to validate the numerical setup.

\subsection{Validation with experiments}\label{subsubsec:validation}

\begin{figure}[hbt]
\centering
\hspace{-2pc}
\begin{subfigure}{0.45\linewidth}
    \includegraphics[height=9pc]{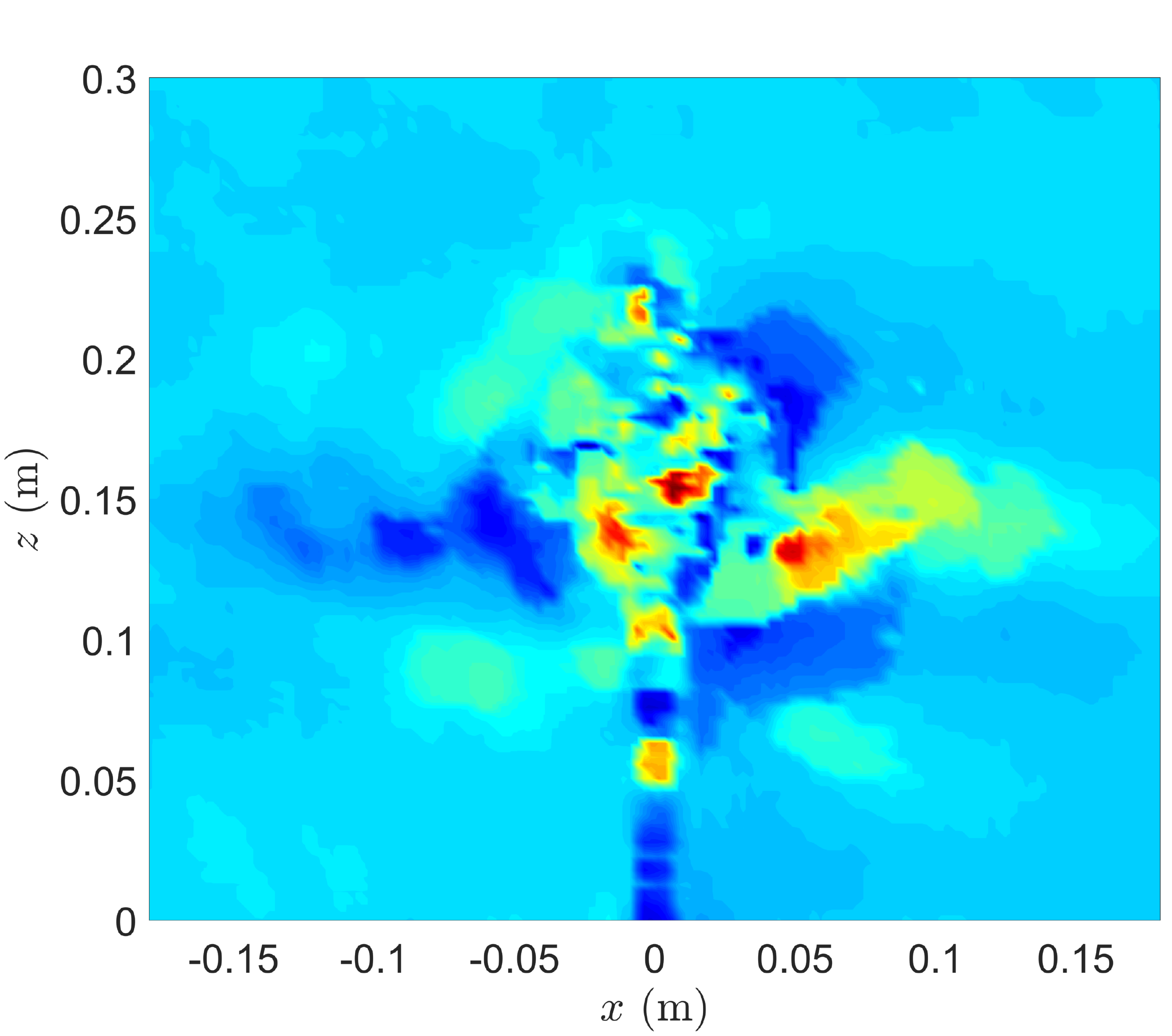}
    \caption{$U_x$ experiment}
    \label{fig:ux_exp}
\end{subfigure}
\hspace{1pc}
\begin{subfigure}{0.45\linewidth}
    \includegraphics[height=9pc]{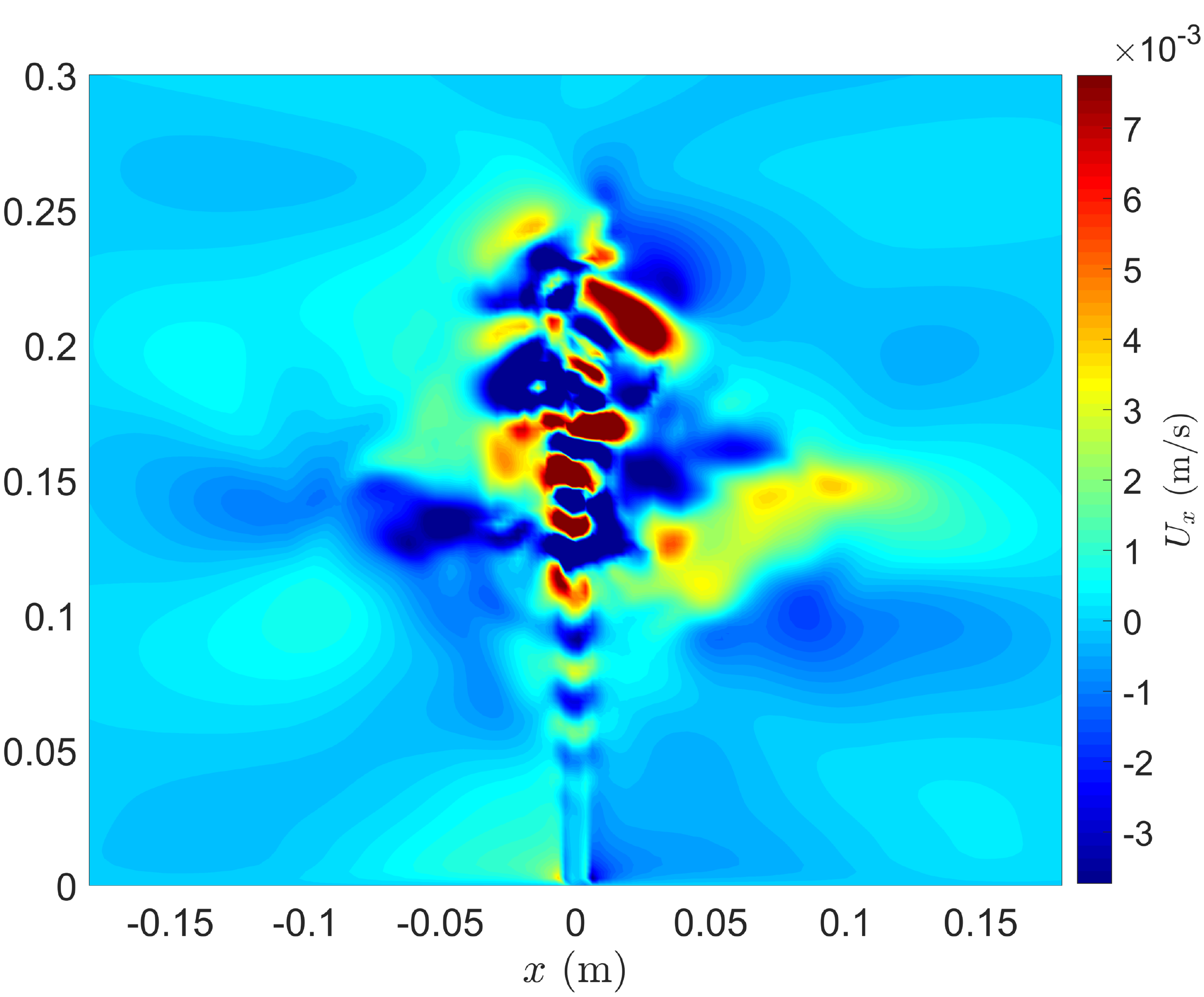}
    \caption{$U_x$ simulation}
    \label{fig:ux_sim}
\end{subfigure}
\\
\hspace{-2pc}
\begin{subfigure}{0.45\linewidth}
    \includegraphics[height=9pc]{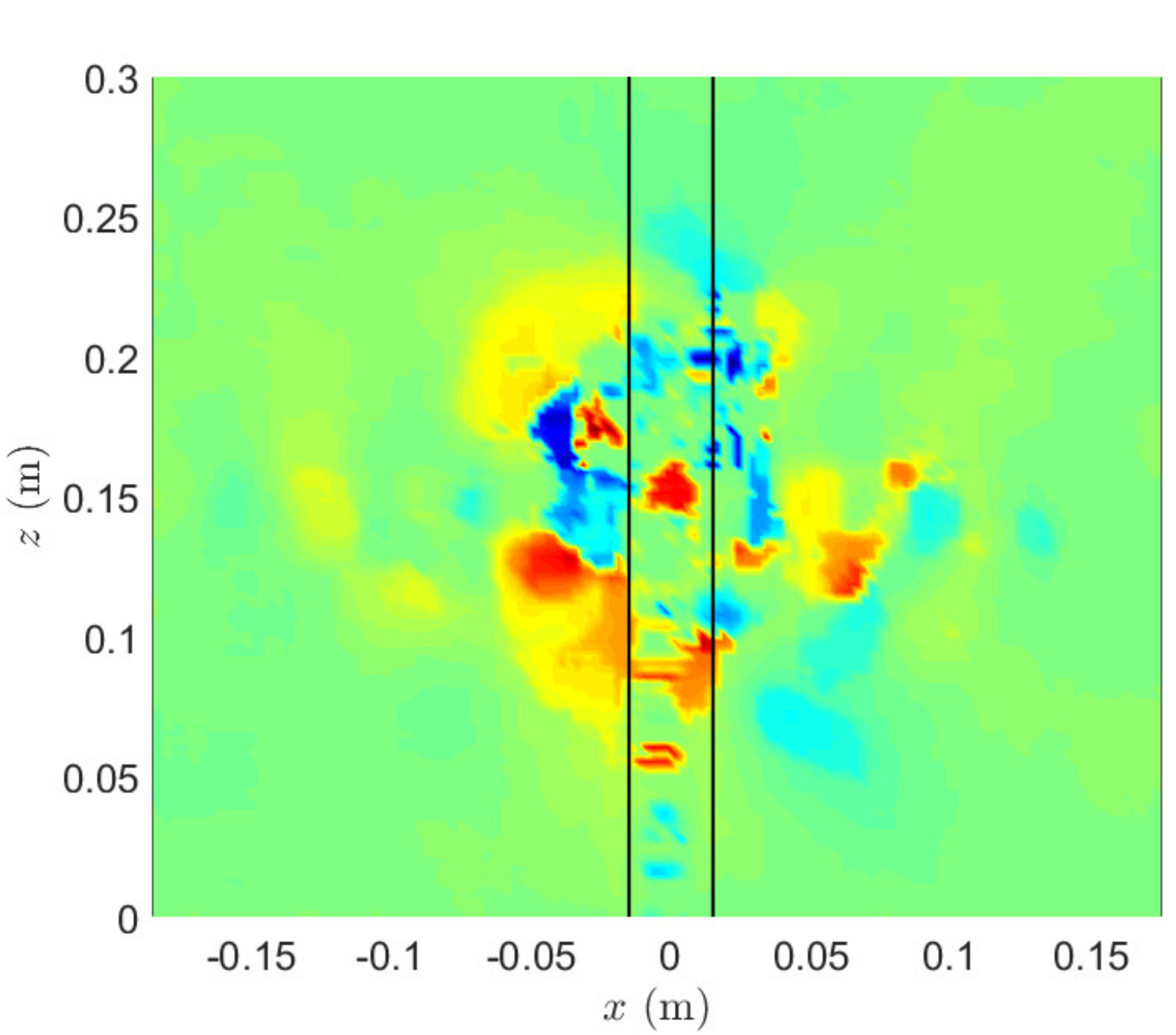}
    \caption{$U_z$ experiment}
    \label{fig:uz_exp}
\end{subfigure}
\hspace{1pc}
\begin{subfigure}{0.45\linewidth}
    \includegraphics[height=9pc]{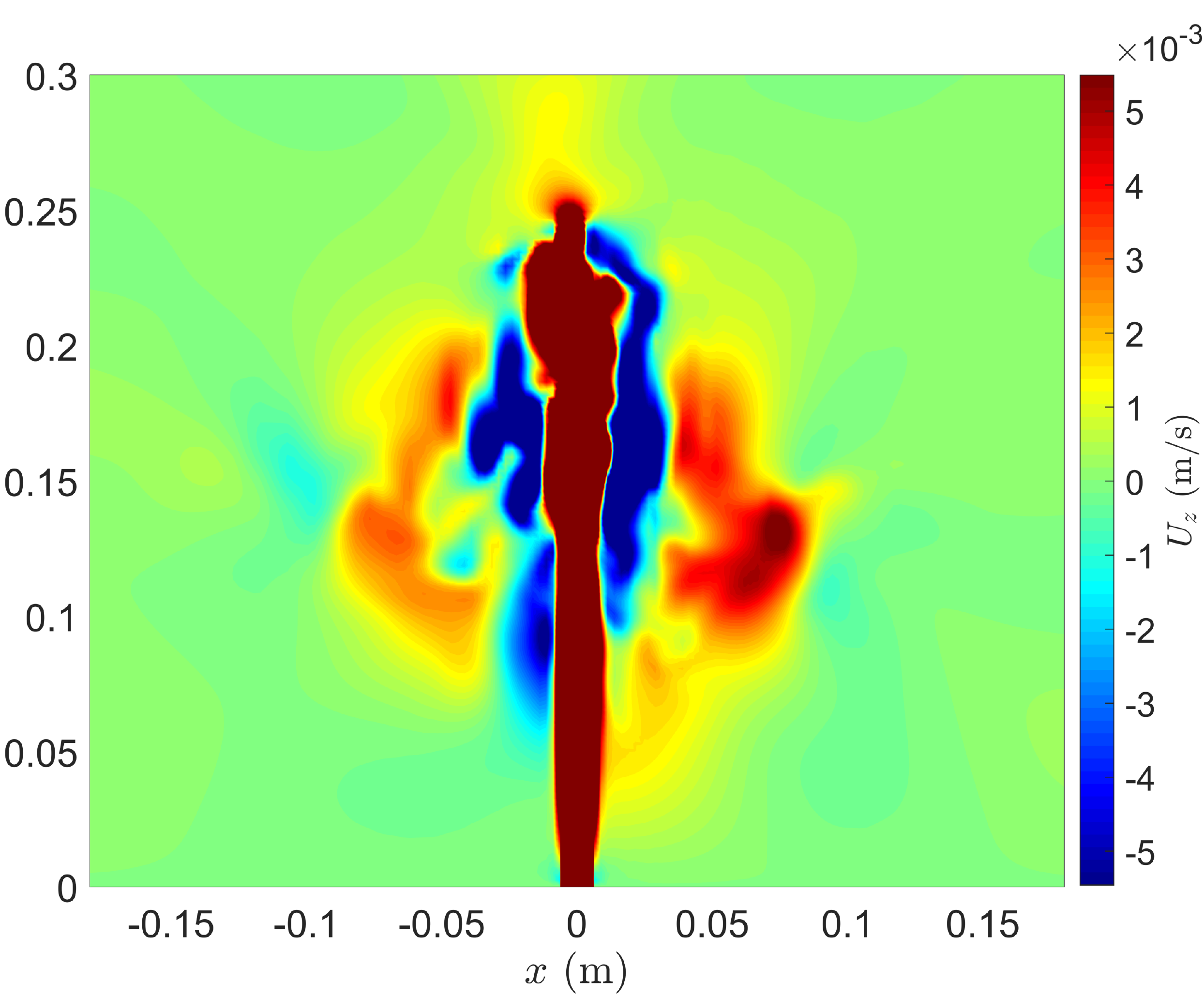}
    \caption{$U_z$ simulation}
    \label{fig:uz_sim}
\end{subfigure}
        
\caption{Horizontal (first row) and vertical (second row) components of the velocity ($U_x$ and $U_z$, respectively) measured in the experiments with the DPIV technique (first column) and from numerical simulations (second column) for the F-15 configuration $t=80$~s over the plane $y=0$. The colour scales are the same within each row. The central jet is not visible in (c) due to the limitations of the CMOS digital camera used in the experiment. Thus, the comparison between (c) and (d) must be made outside the inner jet, i.e., outside the zone delimited with the two vertical black lines in (c).
}
\label{fig:validation_DPIV}
\end{figure}

\begin{figure}[hbt]
\centering
\includegraphics[width=1.0\linewidth]{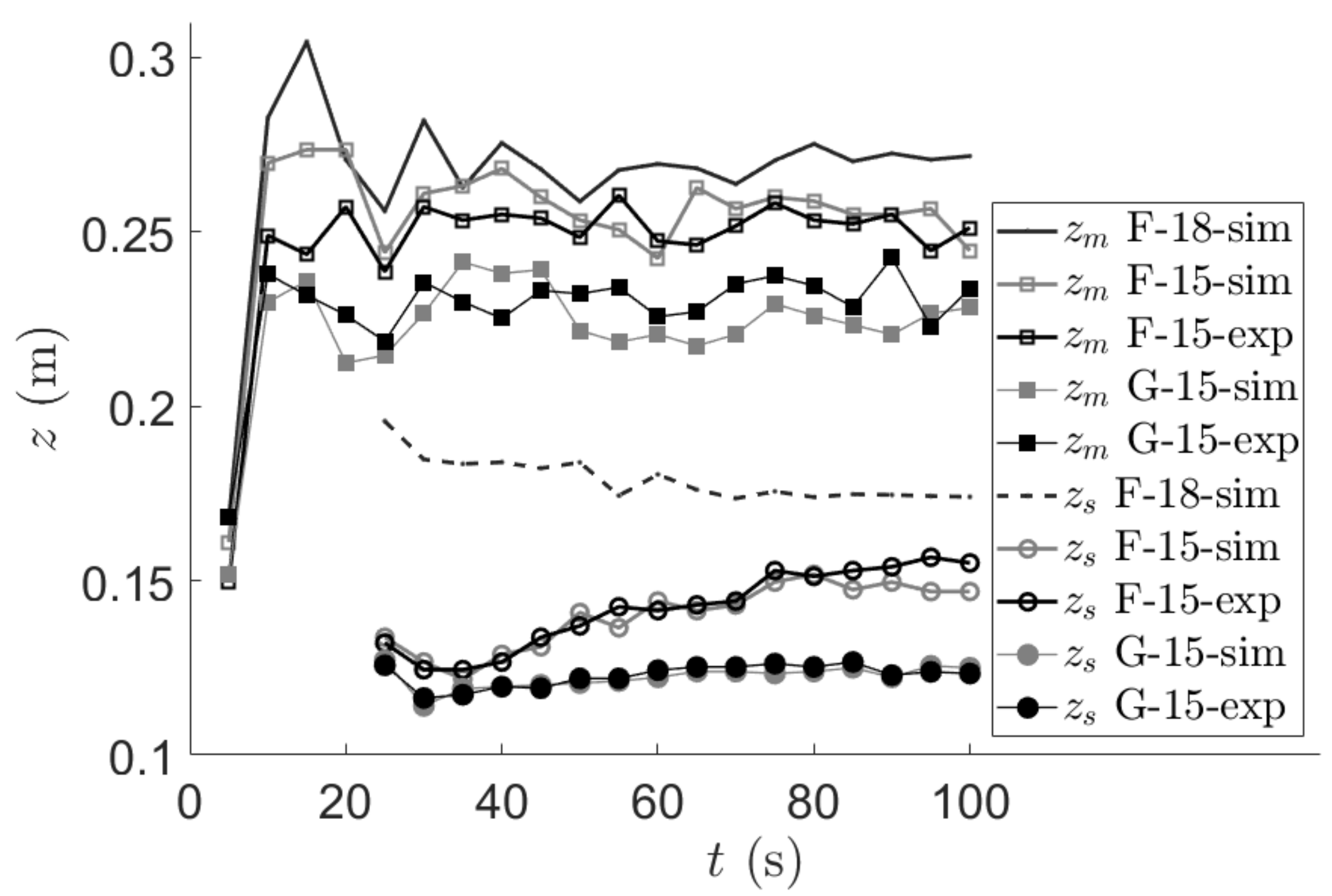}
\caption{
Comparison between the values of $z_m$ (squares) and $z_{s}$ (circles) as functions of time obtained with the experiments\cite{freire2010effect} (black) and numerical simulations
(grey) for the F-15 (empty markers) and G-15 (filled markers) configurations. The solid (dashed) line with no markers corresponds to $z_m$ ($z_{s}$) for the simulations of the F-18 configuration (no experiments were performed in this case).
}
\label{fig:validation_heights}
\end{figure}

In Fig.~\ref{fig:validation_ink}, we compare the numerical and experimental results, showing a good agreement. Moreover, we found the experimental measurements of the 2D velocity fields to be reasonably similar to the simulations, as we illustrate in Fig.~\ref{fig:validation_DPIV}. Although these comparisons correspond to a particular time, the experimental and numerical results for the characteristic heights of the flow, $z_m$ and $z_{s}$, are in good agreement during the whole experiment (about 2~min) for the G-15 and F-15 configurations, as shown in Fig.~\ref{fig:validation_heights}.

\section{\label{sec:results} Results}
\subsection{Particles trajectories}

\begin{figure}[hbt]
\centering
\includegraphics[width=1\linewidth]{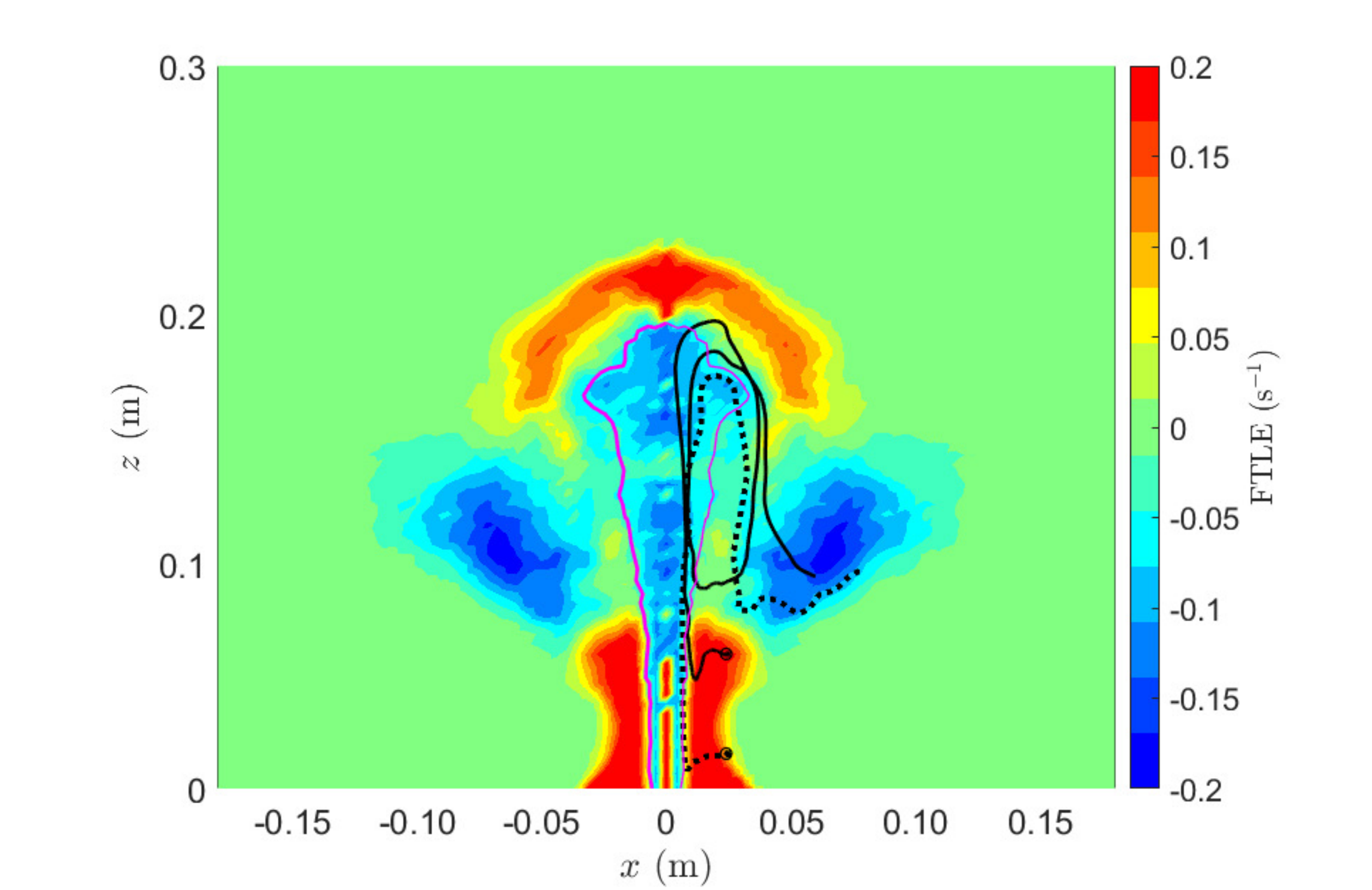}
\caption{
Two particle trajectories (solid and dotted black lines) overlapped with the FTLE field for the G-15 configuration at $t=100$~s. The DAB contour (see Sec.~\ref{sec:calculo_de_entrainment}) is represented with the solid magenta line.}
\label{fig:trayectorias_sobre_FTLE_G15}
\end{figure}

From the  trajectories of the particles initially located  at different places of the environment we obtained the FTLE which is useful to reveal the flow organization of the fountain. In Fig.~\ref{fig:trayectorias_sobre_FTLE_G15} we show some lines of flow of fluid particles initially located near the bottom and outside the jet for the G-15 configuration and  $t\in(0,100$~s$)$. We overlapped some trajectories and the fountain contour (obtained from the contour of the tracer field at a low-level threshold) with the FTLE field. We observe in this figure that the particles initially located near the central jet enter the fountain due to the entrainment process. Most of these particles first rise accompanying the jet and then fall to the spreading front after reaching a maximum height. Then, some particles enter the central jet again, showing that, as assumed in specific models described in the introduction, there is also entrainment between the downflow and the rising jet (see also the supplementary material). We can also observe that in the central region,  the fluid particles follow a prominent upward stream. After reaching a maximum height they descend, and then  rise again to approach the quasi-stationary level.
\subsection{Elucidating the flow organization from the FTLE field} \label{sec:flow_organization}

\begin{figure}[hbt]
\centering
\begin{subfigure}{0.99\linewidth}
    \includegraphics[width=\linewidth]{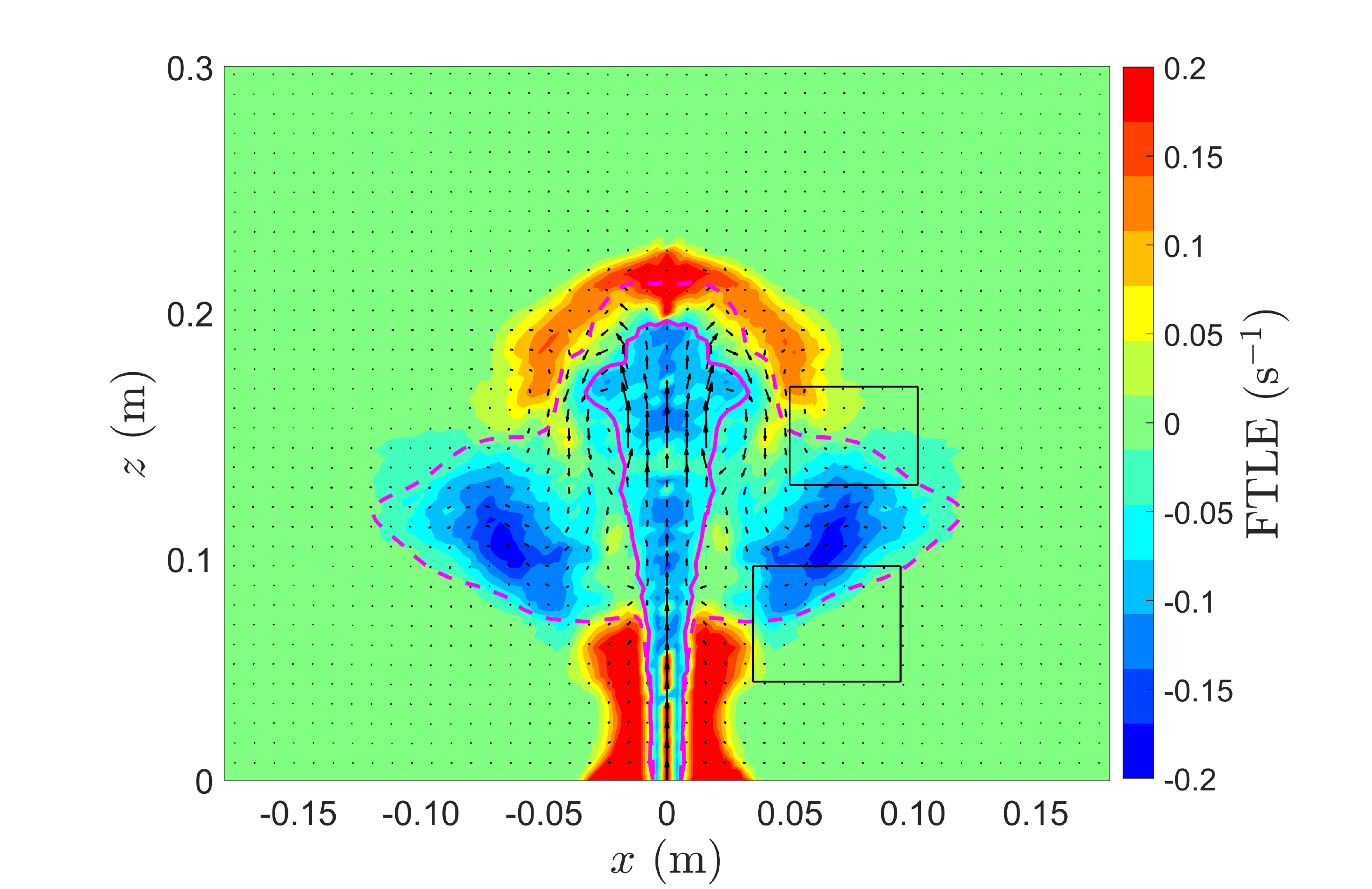}
    \caption{
    FTLE for the G-15 configurations.
    The boxes indicate the  LCS $R_3$ (top) and $R_4$ (bottom) (see  Figs.~\ref{fig:FTLE_G15_zoom_R3} and \ref{fig:FTLE_G15_zoom_R4} for enlarged views).
   }
   \label{fig:FTLE_con_contornos_G15}
\end{subfigure}
\\
\begin{subfigure}{0.990\linewidth}
    \includegraphics[width=\linewidth]{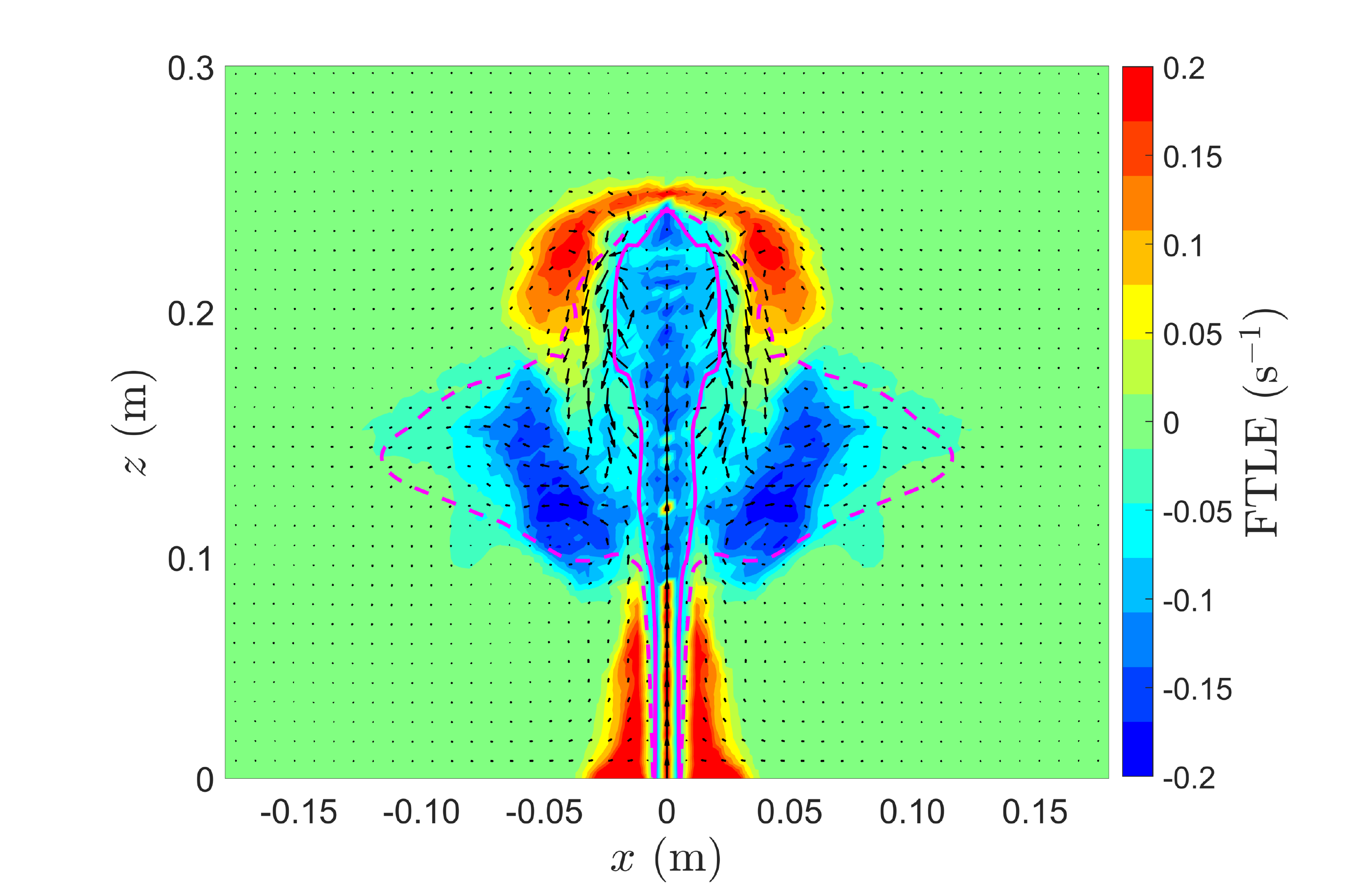}
    \caption{FTLE for the F-15 configuration.}
    \label{fig:FTLE_con_contornos_F15}
\end{subfigure}
\\
\begin{subfigure}{0.90\linewidth}
    \includegraphics[width=\linewidth]{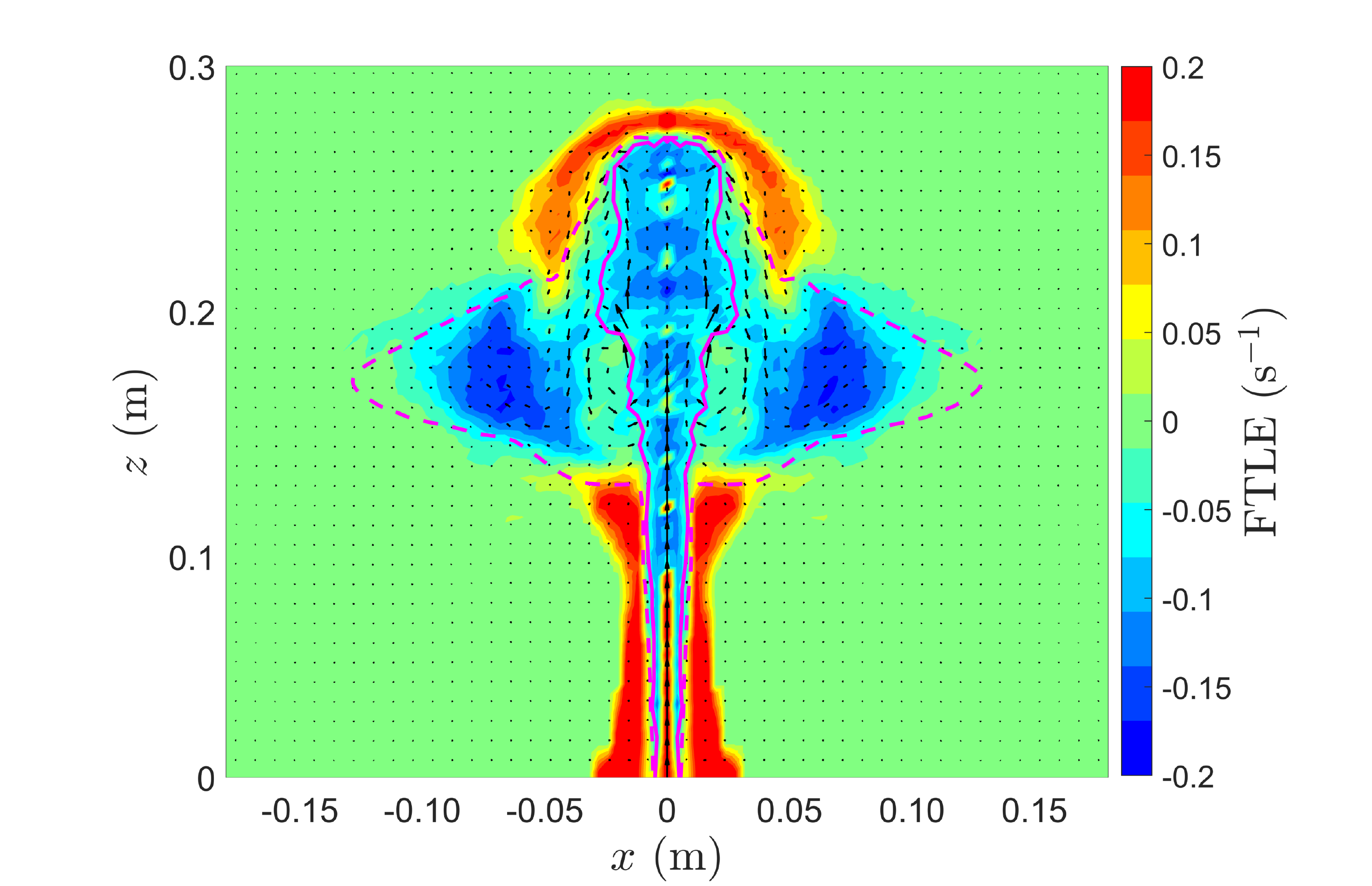}
    \caption{FTLE for the F-18 configuration.}
    \label{fig:FTLE_con_contornos_F18}
\end{subfigure}
\caption{
FTLE field for the three configurations discussed.
The fountain boundary and the DAB  contour are indicated in dashed and solid magenta lines respectively, and the velocity field is represented with black arrows.
}
\label{fig:FTLE_con_contornos}
\end{figure}

In Fig.~\ref{fig:FTLE_con_contornos}, we show the LCS obtained, overlapped with the filtered velocity field (for the sake of clarity only a fraction of the arrows are represented) for a  fully developed flow ($t=100$~s) and  the three configurations considered. It can be seen that the fountain's structure is very similar in the three cases considered.

\begin{figure}[hbt]
\centering
\begin{subfigure}{0.93\linewidth}
    \includegraphics[width=\linewidth]{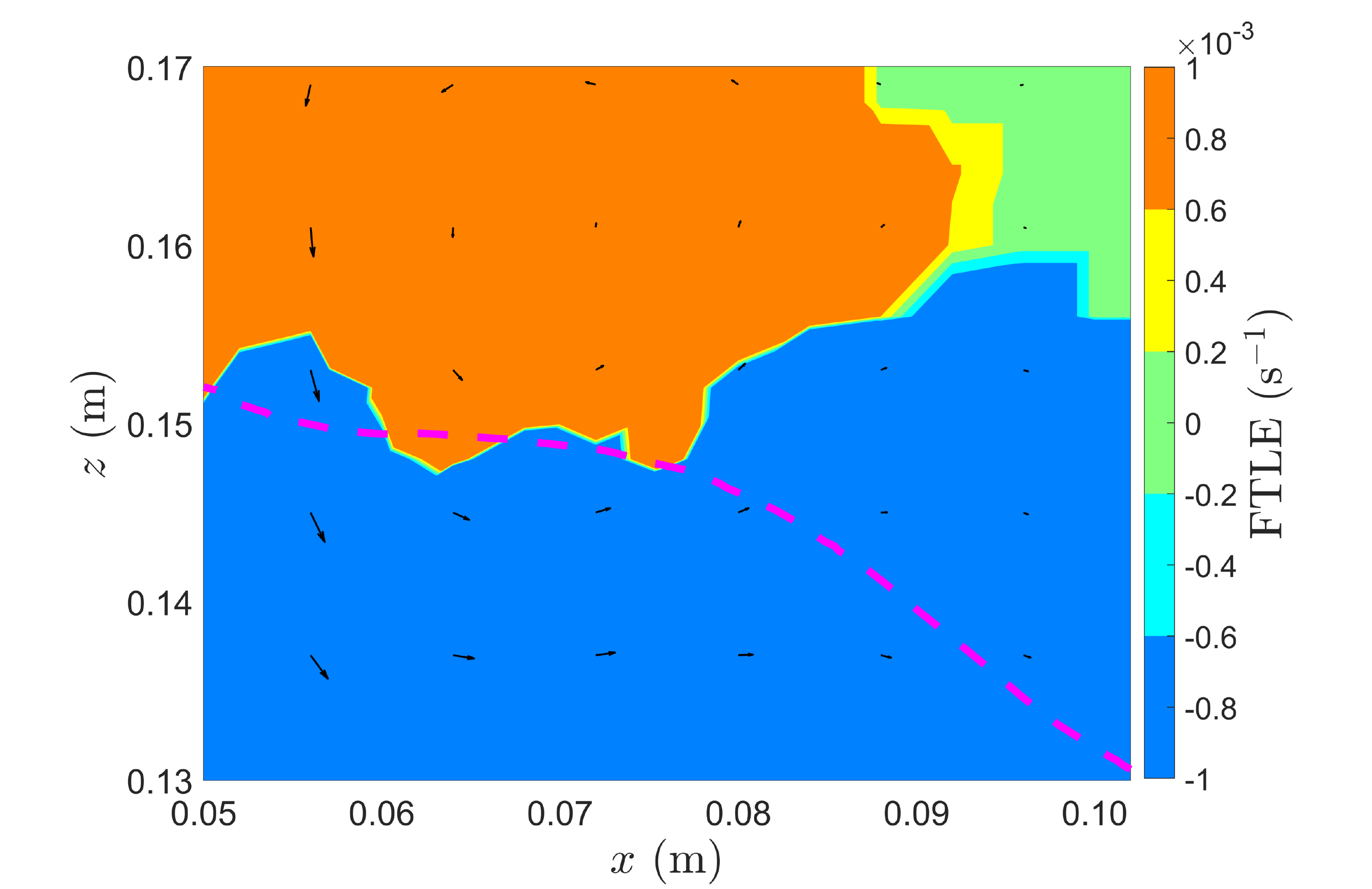}
    \caption{
    Region containing the LCS $R_3$.
    }
    \label{fig:FTLE_G15_zoom_R3}
\end{subfigure}
\\
\begin{subfigure}{0.93\linewidth}
    \includegraphics[width=\linewidth]{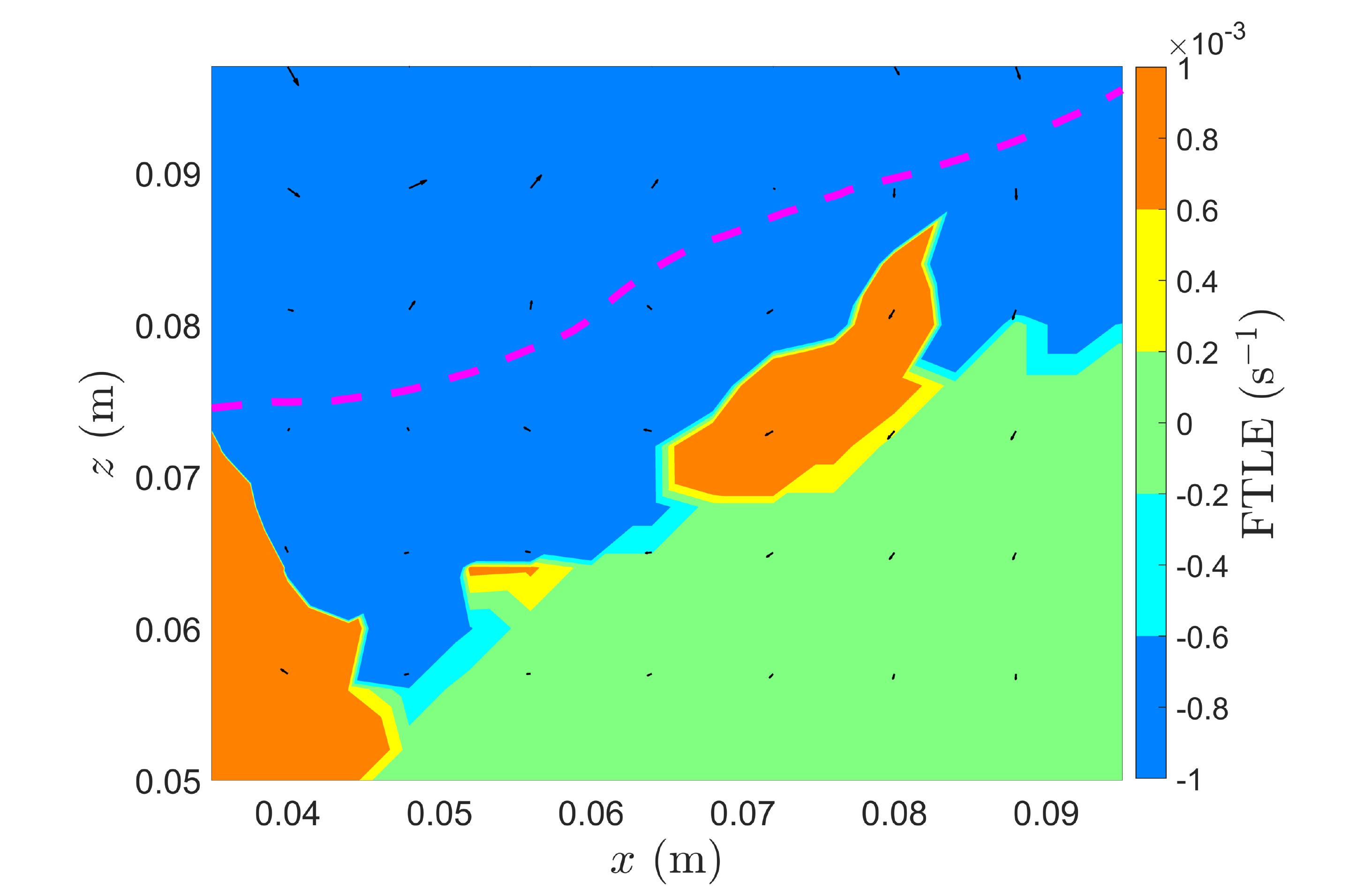}
    \caption{Region containing the LCS $R_4$.}
    \label{fig:FTLE_G15_zoom_R4}
\end{subfigure}
\caption{Enlarged pictures of the two boxes highlighted in Fig.~\ref{fig:FTLE_con_contornos_G15}.}
\label{fig:FTLE_G15_zoom_R3_y_R4}
\end{figure}

\begin{figure}[hbt]
\centering
    \includegraphics[width=0.70\linewidth]{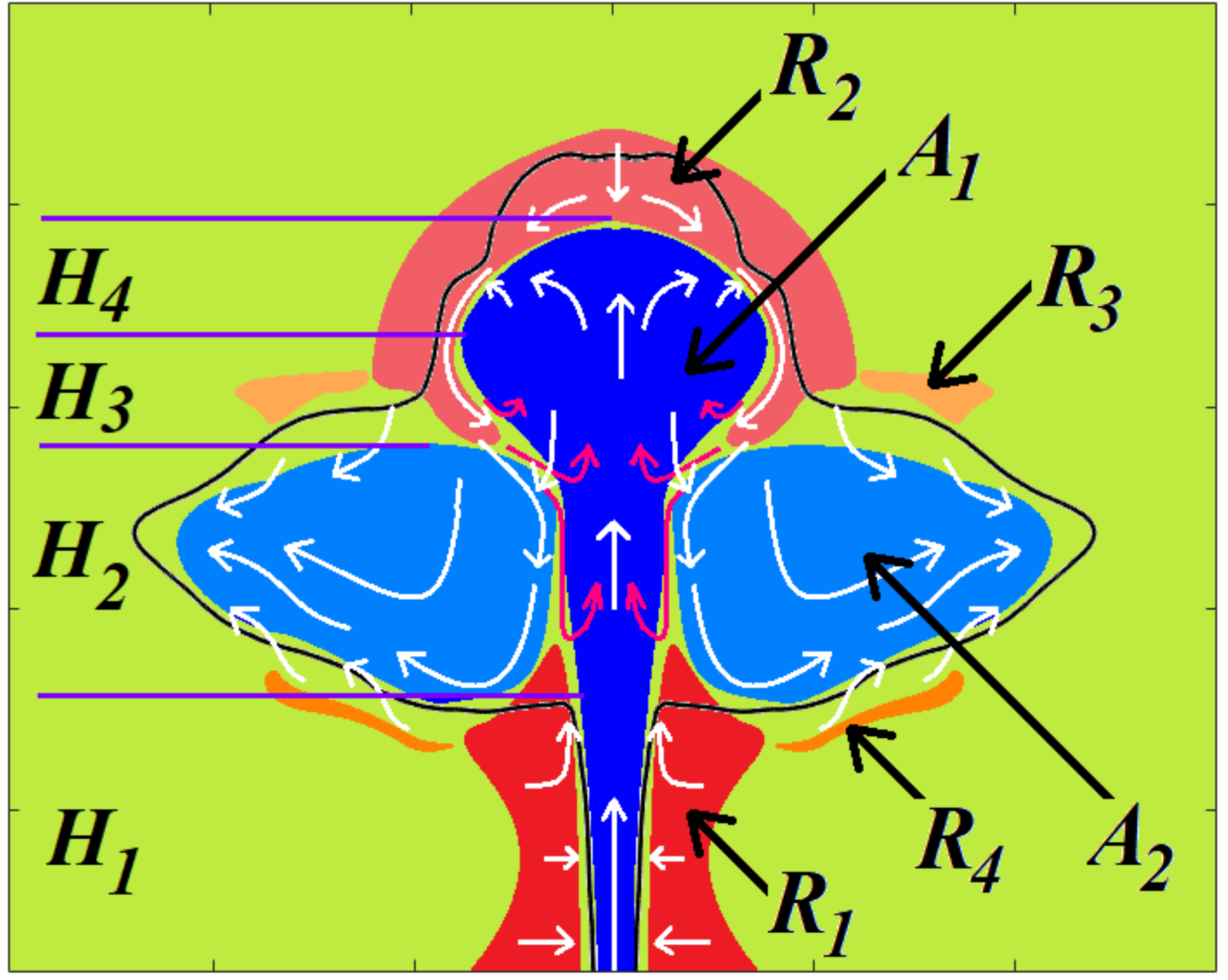}
    \caption{Outline of the LCSs present  in the fountain flow  at $t=100$~s:
    the dominant attractors (repellers) are  $A_1$ and $A_2$ ($R_1$ and $R_2$) indicated in blue (red). The solid black line corresponds to the fountain contour and the arrows indicate the direction of the fluid particles detected in the turbulent fountain (red arrows indicate re-entrainment of fluid into the uprising fountain). We show the limits of the four regions, $H_1$ through $H_4$, where entrainment change its behaviour.
    }
    \label{fig:distribucion_regiones_FTLE}
\end{figure}

By inspecting Fig.~\ref{fig:FTLE_con_contornos}, we distinguish four LCSs denoted as $H_1$ through $H_4$ (outlined in  Fig~\ref{fig:distribucion_regiones_FTLE}), in which we observe qualitative changes in the flow behaviour. The flow features in these  regions are the following:
\begin{itemize}
    \item [$H_1:$]
    A toroidal vertical repeller (LCS $R_1$) surrounds a vertical cylindrical attracting structure along the fountain axis (LCS $A_1$). This attractor-repeller LCS pair produces an FTLE-dipole that pushes the ambient fluid towards the fountain in the neighbourhood of the jet, i.e., the entrainment flow.
    
    \item [$H_2:$]
    in this region, the radial spreading flow takes place. We observe a horizontal toroidal attractor around the jet axis (LCS $A_2$), pulling the fluid away from $r=0$ to join the spreading flow. The spreading \textit{cloud} (see Fig.~\ref{fig:tinta_com}) is delimited by two repeller LCSs, $R_3$ (from the top) and $R_4$ (from the bottom), which are shown in Fig.~\ref{fig:FTLE_G15_zoom_R3_y_R4} since they are significantly weaker than the other LCSs observed. Additionally, within this region, another vertical cylindrical dominant attractor (LCS $A_1$) persists along the fountain axis and re-entrains part of the downflow of fluid particles falling around the rising jet after reaching $z_m$. Finally, the descending fluid particles that remain join the spreading flow through the LCS $A_2$ mentioned first.
    
    \item [$H_3:$]
    the radius of the dominant attractor along the $z$-axis (LCS $A_1$) grows. Ascending fluid particles, heading for $z_m$, come close to reverse velocity and then fall, moving away from this region at its border due to continuity. Depending on the configuration, it may also involve fluid re-entrainment (in our simulations, it only happens for the G-15 configuration).
    
    \item [$H_4:$]
    the top of the central attracting structure (LCS $A_1$) gets weak as it gets closer to $z_m$. Therefore, the main flow points outwards to the $z$-axis at the top of this structure. However, due to a strong repelling structure located on top of it (LCS $R_2$), the fluid reverses direction and falls aside the fountain, heading to regions $H_3$ and $H_2$.
\end{itemize}

In Fig.~\ref{fig:distribucion_regiones_FTLE}, we summarize all the aforementioned LCSs and indicate the fluid's bulk movement with white and red arrows (the lasts correspond to re-entrainment flow). Tab.~\ref{tab:alturas_regiones} notes the dimensionless vertical limits of each region, $z^{*} =z/D$ where $D$ is the nozzle inlet diameter. As expected from Fig.~\ref{fig:validation_heights}, each region in the G-15 configuration is positioned at a lower height compared to the F-15 configuration. The same happens regarding the F-15 characteristic heights, positioned below those corresponding to the F-18 configuration.

\begin{table}[hbt]
\caption{Dimensionless height limits ($z^*$) of each characteristic region for the three configurations.}
\begin{tabular}{c|c|c|c|c}
Configuration & $H_1$    & $H_2$       & $H_3$     & $H_4$         \\ \hline
G-15          & $0 - 9$  & $9 -16$      & $16 - 18$   & $18 - 25$       \\ \hline
F-15          & $0 - 11$   & $11 - 19$     & $19 - 23.5$ & $23.5 - 29.5$ \\ \hline
F-18          & $0 - 16.5$ & $16.5 - 23.5$ & $23.5 - 27$ & $27 - 33.5$
\end{tabular}
\label{tab:alturas_regiones}
\end{table}

We observe from Fig.~\ref{fig:trayectorias_sobre_FTLE_G15} that the fluid particles moving along the upward jet stay within attractor $A_1$ until reaching $z_m$ (it is similar for the three configurations). After that, due to buoyancy and viscous forces, they stop and reverse direction. Due to continuity, they fall at the sides of the rising fountain, escaping from $A_1$. Part of the falling particles moving downwards, aside from $A_1$, re-entrain the upward jet. The rest of the falling fluid joins the spreading flow within region $H_2$, where the spreading \textit{cloud} (as a reference to the shape of the spreading flow from Fig.~\ref{fig:validation_ink}) and the toroidal attracting manifold $A_2$ are located. In summary, we observe pure entrainment over region $H_1$ and re-entrainment over $H_2$ (for the G-15 configuration, re-entrainment is also observed within $H_3$), which is supported by the trajectories of the particles in Fig.~\ref{fig:trayectorias_sobre_FTLE_G15}. Therefore, the boundary surface of $A_1$ is a region of exchange of fluid particles between the rising fountain and its surroundings.

\begin{figure}[htb]
	\centering
	\includegraphics[width=0.97\linewidth,keepaspectratio]{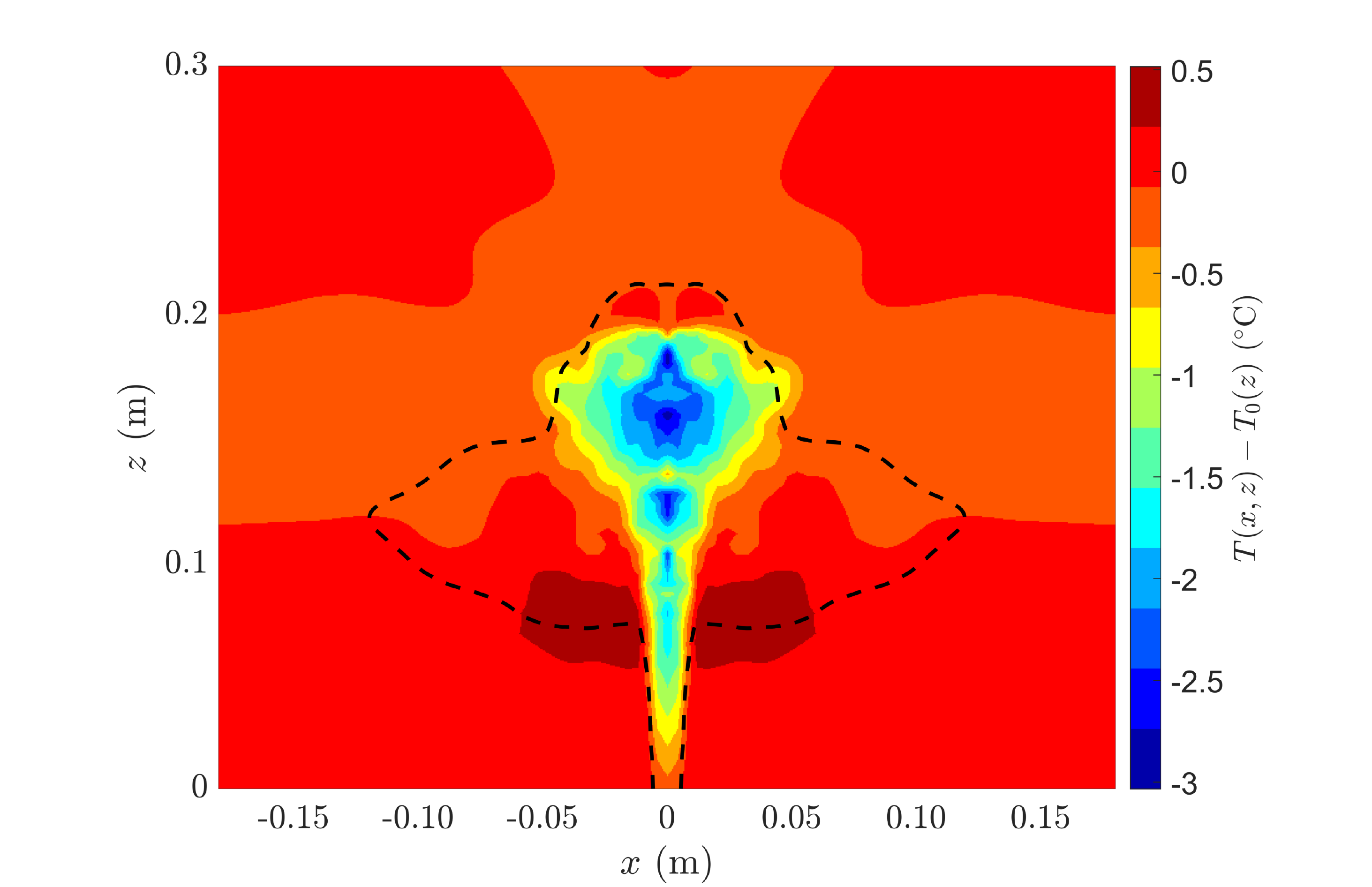}
	\caption{
	Buoyancy force at time 100~s for the G-15 configuration. At the bottom of the spreading \textit{cloud} (within a radius of 0.06~cm and a height between approximately 6 and 9~cm), the buoyancy force points upwards. Such force prevents the fluid from falling beyond the spreading height, i.e., it originates the LCS $R_4$.
	}
	\label{fig:buoyancy_100s_Ts1995_grid}
\end{figure}

The spreading flow within region $H_2$ is a consequence of the temperature difference between the fountain, $T(x,z)$, and the stratified environment at that level, $T_0(z)$, as seen in Fig.~\ref{fig:buoyancy_100s_Ts1995_grid}. More precisely, since $ T(x,z)-T_0(z)>0 $ at the bottom of this region, the buoyant force points upwards, and therefore, once the downflow reaches this point, it is forced to stop and rise again, entering the spreading flow due to continuity. Depending on the configuration (see Tab.~\ref{tab:caudales_regiones}), part of it may also add to the re-entrainment.

The gravity (buoyancy) force field highlights the physical mechanism behind the emergence of the main LCS. For instance, around the top limit of region $H_1$ and $0.02<r (\mathrm{m})<0.06$, the attracting LCS defines a region where particles agglomerate, keeping their height and not falling again. This behaviour correlates an $R_4$ repelling LCS, weak in comparison with $R_1$ and $R_2$ (see Fig.~\ref{fig:FTLE_con_contornos_G15}), to the region of positive buoyancy of Fig.~\ref{fig:buoyancy_100s_Ts1995_grid} where the spreading \textit{cloud} stands (i.e., region $H_2$). The $A_1$ LCS, extending from $H_1$ to $H_3$, correlates to a negative buoyancy region from Fig.~\ref{fig:buoyancy_100s_Ts1995_grid} (where entrainment and re-entrainment takes place). There also exists another weak repelling LCS, identified as $R_3$ in Fig.~\ref{fig:distribucion_regiones_FTLE}, on the top of the $A_2$ LCS, which leads the spreading flow to join the spreading \textit{cloud} and prevent it from moving upwards ($R_3$ is shown in Fig.~\ref{fig:FTLE_con_contornos_G15}).

In summary, once the flow is fully developed, i.e. it is steady far from the fountain axis (since the turbulent fluctuation can only be strong near the axis), the mixing occurs on account of two agents: 1) the entrainment of ambient fluid to the ascending jet (within region $H_1$), and 2) the re-entrainment of part of the fluid that descends enclosing the ascending fountain (mainly in region $H_2$, and depending on the configuration, also present in $H_3$), promoting a shear flow where mixing takes place. As we discuss in the next section, the in depth analysis and measurement of such mixing mechanisms, not conducted before to the best of our knowledge, is a notable contribution of this work.

\subsection{Entrainment and re-entrainment quantification} \label{sec:calculo_de_entrainment}

Within structure $A_1$ lies the ridge of the $\Lambda^-$ field, located along the centre of the upward jet, i.e., $r=0$ axis (see Fig.~\ref{fig:FTLE_con_contornos}). We refer to this $\Lambda^-$ ridge as the Dominant Attractor (DA) and, as observed, particles do not cross the DA (see Fig.~\ref{fig:trayectorias_sobre_FTLE_G15}), while those initially close to the DA tend to be attracted to it. Therefore, it can be defined as an influence zone where particles that enter are likely to be attracted towards the DA. By setting a suitable threshold for the corresponding FTLE field (e.g., about $-0.02$ for the G-15 configuration), we define the boundary surface that envelops the dominant attracting LCS along the axis as $0<z<z_m$. We refer to this surface as the Dominant Attractor Boundary (DAB). In Fig.~\ref{fig:trayectorias_sobre_FTLE_G15} we plot the DAB for the G-15 configuration (which appears as contour in the 2D plot). The mentioned FTLE threshold was chosen \textit{ad-hoc} so that, when compared with particles trajectories, the DAB corresponds to a boundary crossed by particles. Consequently, particle trajectories across the DAB are mostly perpendicular to the DAB, making this boundary suitable for computing the entrainment and re-entrainment flows.

By inspecting the particle trajectories from Fig.~\ref{fig:trayectorias_sobre_FTLE_G15} we reveal a pure entrainment mixing within region $H_1$, while re-entrainment and the escape of fluid particles coexist within $H_2\cup H_3$. The same occurs in configurations F-15 and F-18. Moreover, despite the flow's complexity, the velocity direction across the DAB agrees with the described mixing processes, as shown in Fig.~\ref{fig:FTLE_con_contornos}, where the FTLE fields for the G-15 and F-15 configurations and the corresponding DABs and velocity fields are overlapped.

\begin{figure}[htb]
	\centering
	\includegraphics[width=1.0\linewidth]{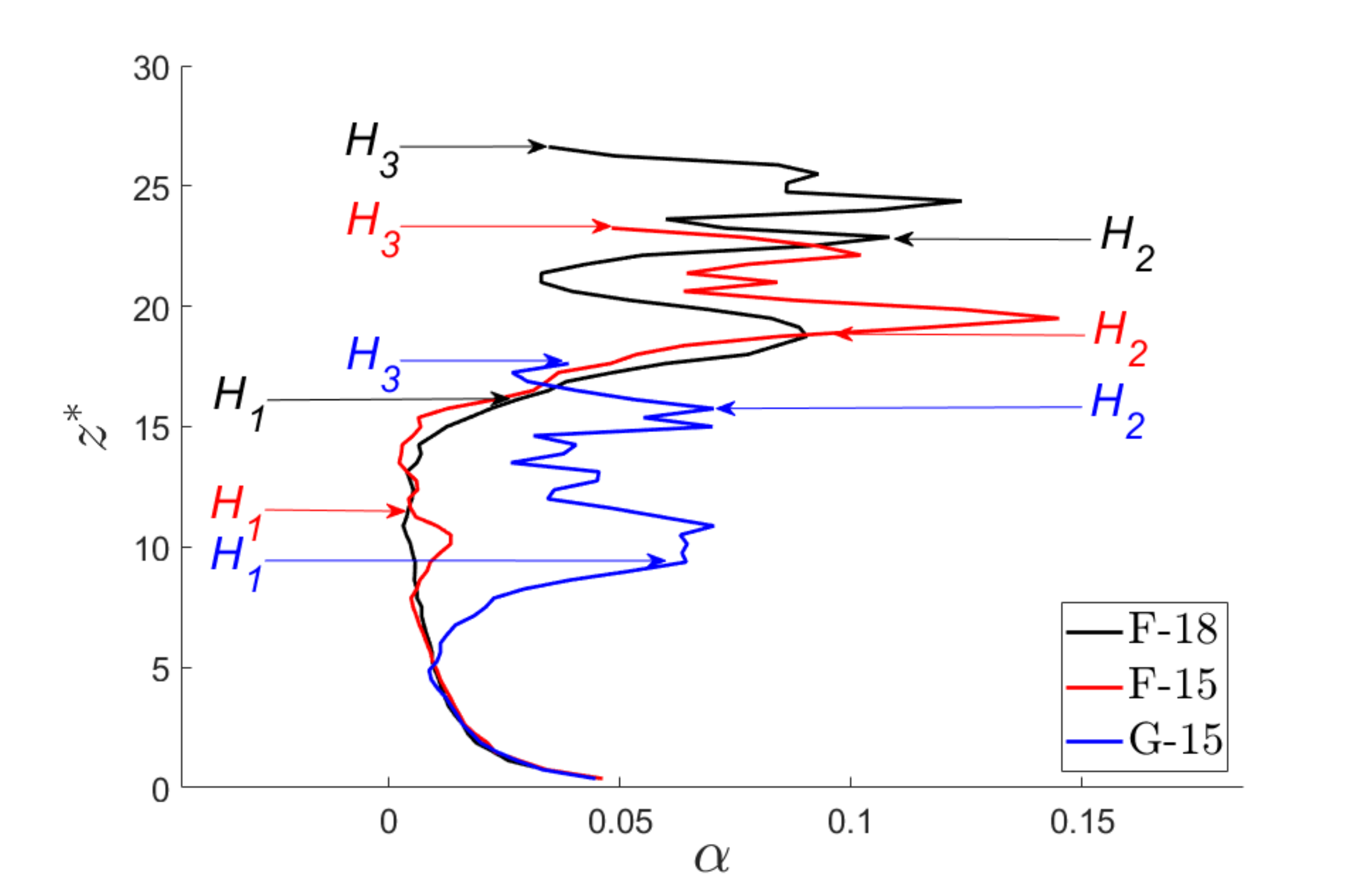}
	\caption{Coefficient $\alpha$ at time $t=100$~s, for the three configurations: G-15 (blue line), F-15 (red line) and F-18 (black line). The limits of the characteristic regions from Tab.~\ref{tab:alturas_regiones} are also indicated by arrows of the same colour code as the $\alpha$ curves.}
	\label{fig:alpha_and_regions}
\end{figure}

As a notable contribution of our work, we computed the entrainment coefficient $\alpha$ from Morton et al.~\cite{morton1956turbulent}, extended to regions above $H_1$ as proposed by Sarasua et al.~\cite{sarasua2021spreading}. First, at each computated height $z$, we calculated the mean vertical velocity in the region within the DAB boundaries, $w_{up}(z)$. Secondly, in regions $H_2$ and higher, we computed the downwards vertical velocity outside the DAB, $w_{down}(z)$ (in region $H_1$ there is pure entrainment, i.e., $w_{down}=0$).
Finally, we interpolated the horizontal velocity into the DAB, $u_{entr}(z)$ and computing $\alpha(z)$ as follows:
\begin{equation}
\displaystyle \alpha (z) = \dfrac{w_{up}(z)-w_{down}(z)}{u_{entr}(z)}
\label{eq:alpha_extended}
\end{equation}
In Fig.~\ref{fig:alpha_and_regions} we show $\alpha(z)$ at time $t=100$~s, for the three configurations. Turbulence plays a crucial role in the entrainment process, as it can be observed from the $\alpha$ profile for the G-15 configuration in Fig.~\ref{fig:alpha_and_regions}, which grows faster than the F-configurations within the region $H_1$. Nevertheless, the overall behaviour of $\alpha$ in the F-configurations is similar, although F-18 entrains ambient fluid from higher strata layers than F-15 since the opposing buoyancy force is smaller for the former. Besides the current analysis, we remark that we measured $\alpha$ for the whole uprising fountain, providing our DAB definition's mixing criteria.

We calculated three relevant parameters to improve the state-of-the-art theoretical models of turbulent fountains from our analysis. Such parameters are the mean entrainment coefficient $\overline{\alpha}(H_1)$ (computed as the average of $\alpha$ within region $H_1$, ), the mean re-entrainment coefficient $\overline{\alpha}(H_2)$ (computed as the average of $\alpha$ over $H_2$), and the parameter $\gamma$, defined in Sarasua et al.~\cite{sarasua2021spreading}, which reads as follows:
\begin{equation}
\gamma = \dfrac{T_0(z_s)-T(z_m)}{\dfrac{T_0(z_m)+T_0(z_s)}{2}-T(z_m)},
\label{eq:gamma_def}
\end{equation}
where $T_0(z)$ is the initial ambient stratification temperature at height $z$ and $T(z_m)$ is the fountain temperature at the maximum height. Note that the parameter $\overline{\alpha}(H_1)$ is the one defined simply as the constant entrainment coefficient $\alpha$ in the original work of Morton et al.~\cite{morton1956turbulent}.

All these parameters are shown in Tab.~\ref{tab:alphas_y_gammas} for the three configurations. The obtained values of $\gamma$ are found in good agreement with the reported by Sarasua et al.~\cite{sarasua2021spreading}. The values of $\alpha$ and $\alpha_r$ are significantly different in every configuration, with $\alpha_r$ always being greater. On the other hand, $\alpha_r$ is similar for the G-15 and F-18 configurations, showing that increasing either the temperature or the turbulence of the fountain favours the re-entrainment mechanism for the considered configurations.

\begin{table}[hbt]
\caption{Main parameters for future refinement of the theoretical model of turbulent fountains: the entrainment and re-entrainment coefficients ($\overline{\alpha}(H_1)$ and $\overline{\alpha}(H_2)$, respectively), and the characteristic downflow mixing parameter ($\gamma$) defined in Sarasua et al.~\cite{sarasua2021spreading}.}
\begin{tabular}{c|c|c|c}
Configuration & $\overline{\alpha}(H_1)$ &
$\alpha_r$ &
$\gamma$ \\ \hline
G-15 & $0.021$ & $0.051$ & $0.25$ \\ \hline
F-15 & $0.014$ & $0.021$ & $0.15$ \\ \hline
F-18 & $0.012$ & $0.062$ & $0.20$
\end{tabular}
\label{tab:alphas_y_gammas}
\end{table}

In addition, we computed the cumulative inflow across the DAB as follows. Given the mesh discretisation introduced for the numerical simulations, the DAB is composed of small line segments, $s_i$, of length $ds_i$, with maximum and minimum distances to the $z$ axis $r^M_i$ and $r^m_i$, respectively, and maximum and minimum heights ($z^M_i$ and $z^m_i$, respectively). Therefore, the flow across the DAB is the sum of the flow over all the elements $s_i$.

\begin{figure}[htb]
	\centering
	\includegraphics[width=1.0\linewidth]{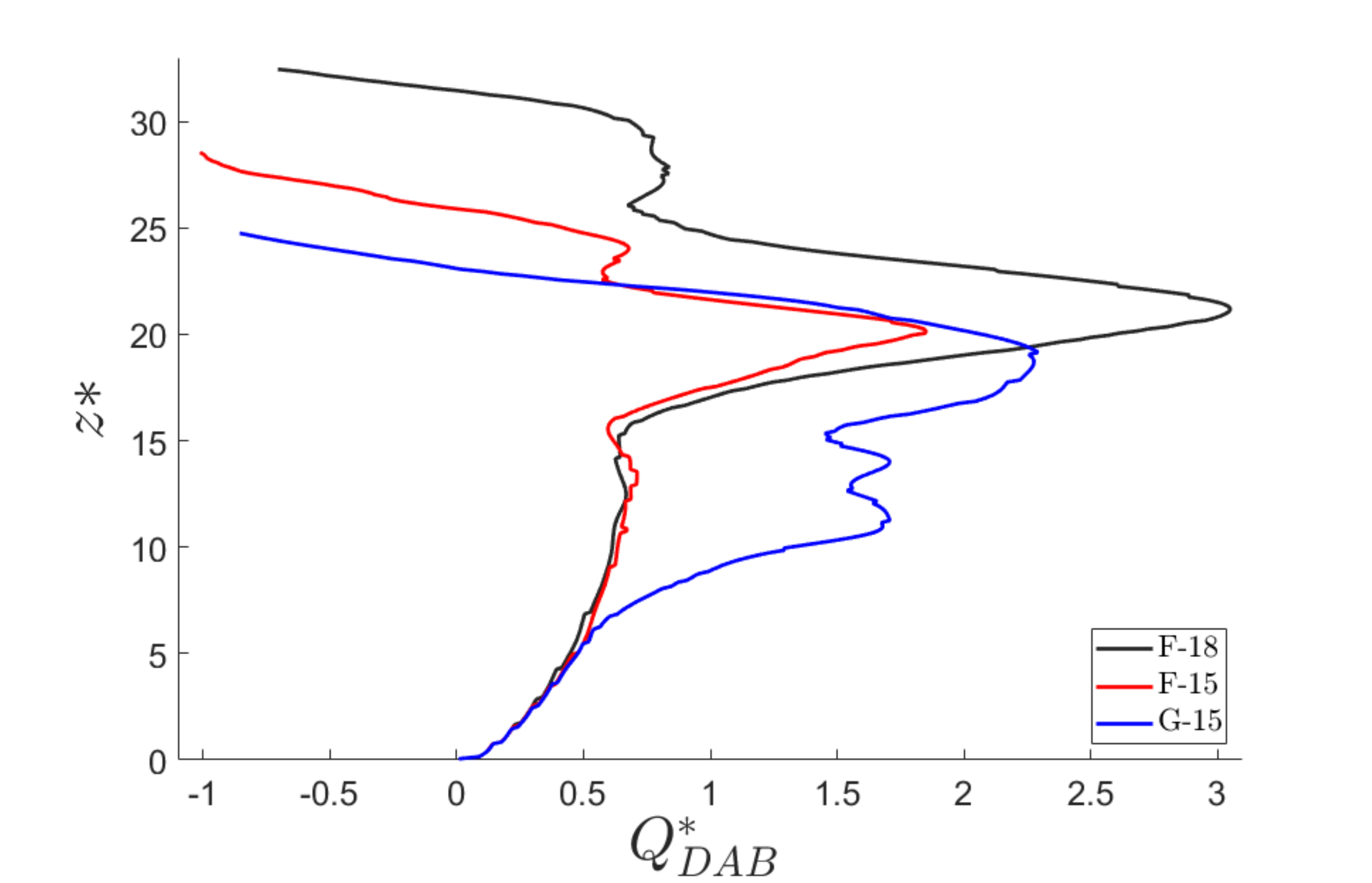}
	\caption{Evolution of the dimensionless cumulative flow through the DAB ($Q^*_{DAB}$) with the dimensionless height $z^*$ ($z^*=z/D$, where $D$ is the inlet nozzle diameter) for the G-15 (blue), F-15 (red) and F-18 (black) configurations. The limits of the four distinct regions, $H_1$ to $H_4$ (Tab.~\ref{tab:alturas_regiones}), are indicated with arrows.}
	\label{fig:cumulative_entrainment}
\end{figure}

We proceeded by compute the unit vector perpendicular to $s_i$ and pointing inwards the DAB, $\hat{n}_i$, and the interpolated velocity vector in the midpoint of $s_i$, $\vec{u}_i$.
Since we assume azimuthal symmetric flow around the $z$ axis (due to the smoothing process), the total inflow across the DAB between $z^m_i$ and $z^M_i$ corresponds to:
\begin{equation}
    dq_{DAB}^i = \vec{u}_i \cdot \hat{n}_i dA_i,
\end{equation}
where $dA_i=\pi \left( r^m+r^M \right) ds_i$ is the surface of a cylinder defined by the rotation of the line segment $ds_i$ around the $z$-axis. We also define the cumulative flow through the DAB at height $z$, $Q_{DAB}(z)$, as follows:
\begin{equation}
Q_{DAB}(z) = \sum_{z^M_i \leq z} q^i_{DAB},
\end{equation}
provided $0<z<z_m$. We also define the non-dimensional cumulative flow through the DAB as $Q^*_{DAB}=Q_{DAB}/q_{in}$, where $q_{in}$ is the inflow rate through the nozzle port. The plots of $Q^*_{DAB}$ for the three different configurations are shown in Fig.~\ref{fig:cumulative_entrainment}. In that figure, we also point to the vertical limits of each region $H_1$ to $H_4$.
In addition, in Tab.~\ref{tab:caudales_regiones} we show the computed values of $Q^*_{DAB}$ within each $H_i$ region, i.e., 
\begin{equation}
\centering
Q^*_{DAB}(H_i)=Q^*_{DAB}(z_i^M)-Q^*_{DAB}(z_{i-1}^M),
\label{eq:cumultative_entH_region}
\end{equation}
for $i=1,\dots,4$, where $z_0^M=0$ and $Q^*_{DAB}(z_0^M)=0$.

\begin{table}[hbt]
\caption{$Q^*_{DAB}$ over each region.}
\begin{tabular}{c|c|c|c|c}
Configuration & $Q^*_{DAB}(H_1)$ & $Q^*_{DAB}(H_2)$ & $Q^*_{DAB}(H_3)$ & $Q^*_{DAB}(H_4)$  \\ \hline
G-15 & 1.0   & 0.67  &  0.53  & -3.1  \\ \hline
F-15 & 0.65  & 0.78  & -0.8   & -1.65 \\ \hline
F-18 & 0.85  & 0.85  & -0.89  & -1.5
\end{tabular}
\label{tab:caudales_regiones}
\end{table}
As expected, the sum of $Q^*_{DAB}$ over all the regions equals $-1$ for each configuration, which corresponds to the drain of the inflow. Tab.~\ref{tab:caudales_regiones} not only shows that G-15 is the most efficient configuration regarding the entrainment of ambient fluid, but also that in the re-entrainment process, since its extension is over regions $H_2$ and $H_3$, whereas in F-configurations re-entrainment only occurs in $H_2$. This observation is in agreement with Fig.~\ref{fig:FTLE_con_contornos_F15}, in the sense that laminar flow continues more orderly than turbulent flows at higher heights, so the top of the DAB is expected to be more compact and large in height for the F-configurations, thus preventing radial flows (re-entrainment) until the downflow enters lower layers (region $H_2$).

\subsection{Efficiency in the withdrawal of harmful environmental fluid.}

A remarkable technological application of turbulent fountains is the Selective Inverted Sink~\cite{frostprotection_url,guarga2000evaluation,freire2010effect} (SIS), mainly devoted to mitigate frost damage in agriculture under radiation frost conditions, among other featured applications~\cite{frostprotection_url}. In this scenario, the surface temperature inversion occurs, i.e., a stable air stratification where the dense cold lays on the ground surface and may reach harmful temperatures for the crops. The SIS consists of a chimney-like structure whose selective action withdraws that lower cold air strata, resulting in a turbulent fountain flow of the ejected air.

Improving the device's efficiency requires a deep understanding of the developed flow structures, depending on the ejected flow conditions, like turbulence and temperature. In this section, we compute the removed volume of the lower strata of ambient fluid based on the results from the previous sections. On the one hand, the ejected flow $q_{in}$ is due to the ejected air by the SIS. On the other hand, 
the uprising fountain captures air from its surroundings within region $H_1$, i.e., the pure entrainment mechanism.

\begin{figure}[hbt]
\hspace{-1pc}
\centering
\begin{subfigure}{0.4\linewidth}
    \includegraphics[height=16pc]{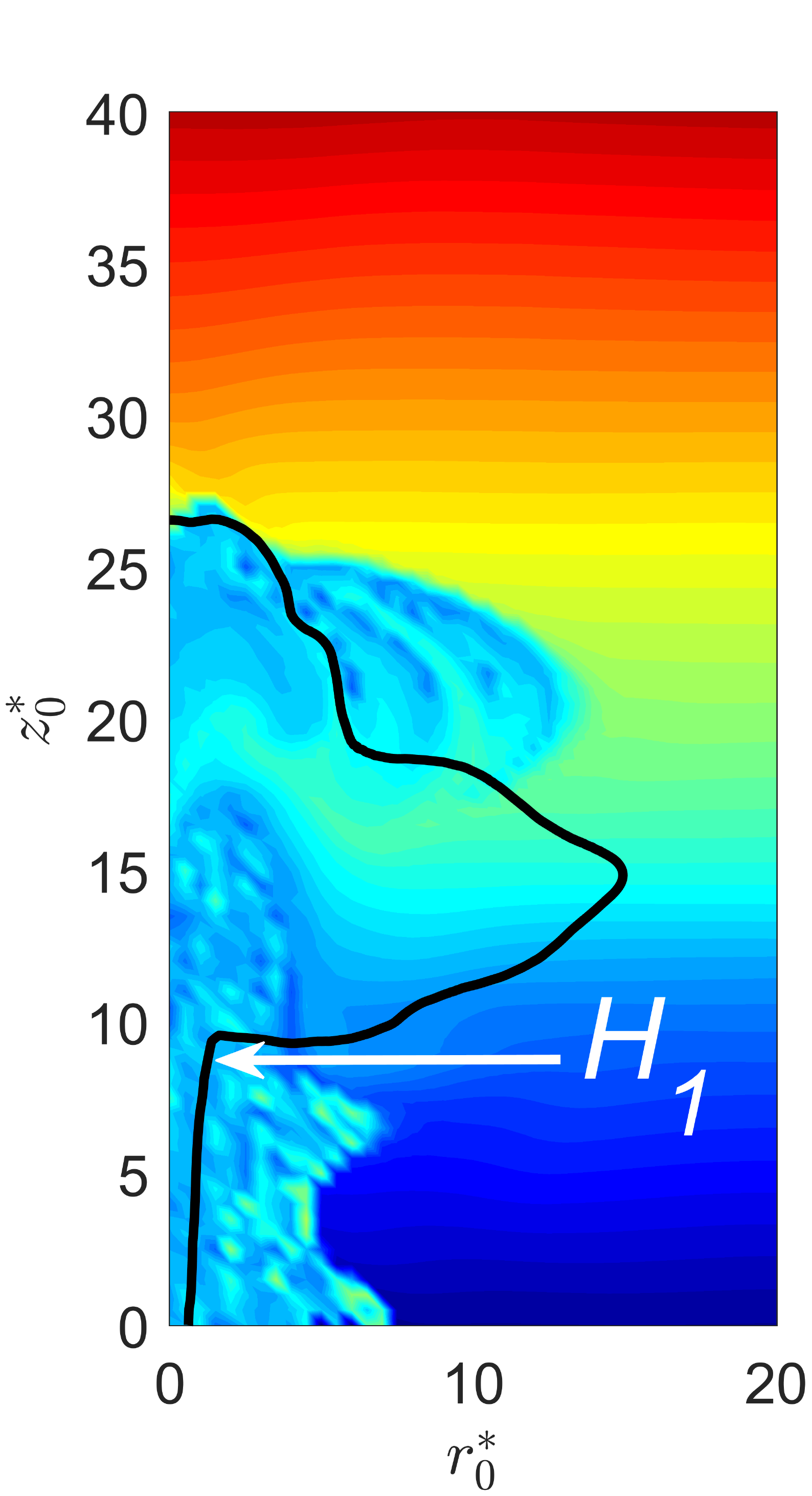}
    \caption{G-15 configuration.}
    \label{fig:zf_con_contornos_G15}
\end{subfigure}
\hspace{0.1pc}
\begin{subfigure}{0.6\linewidth}
    \includegraphics[height=16pc]{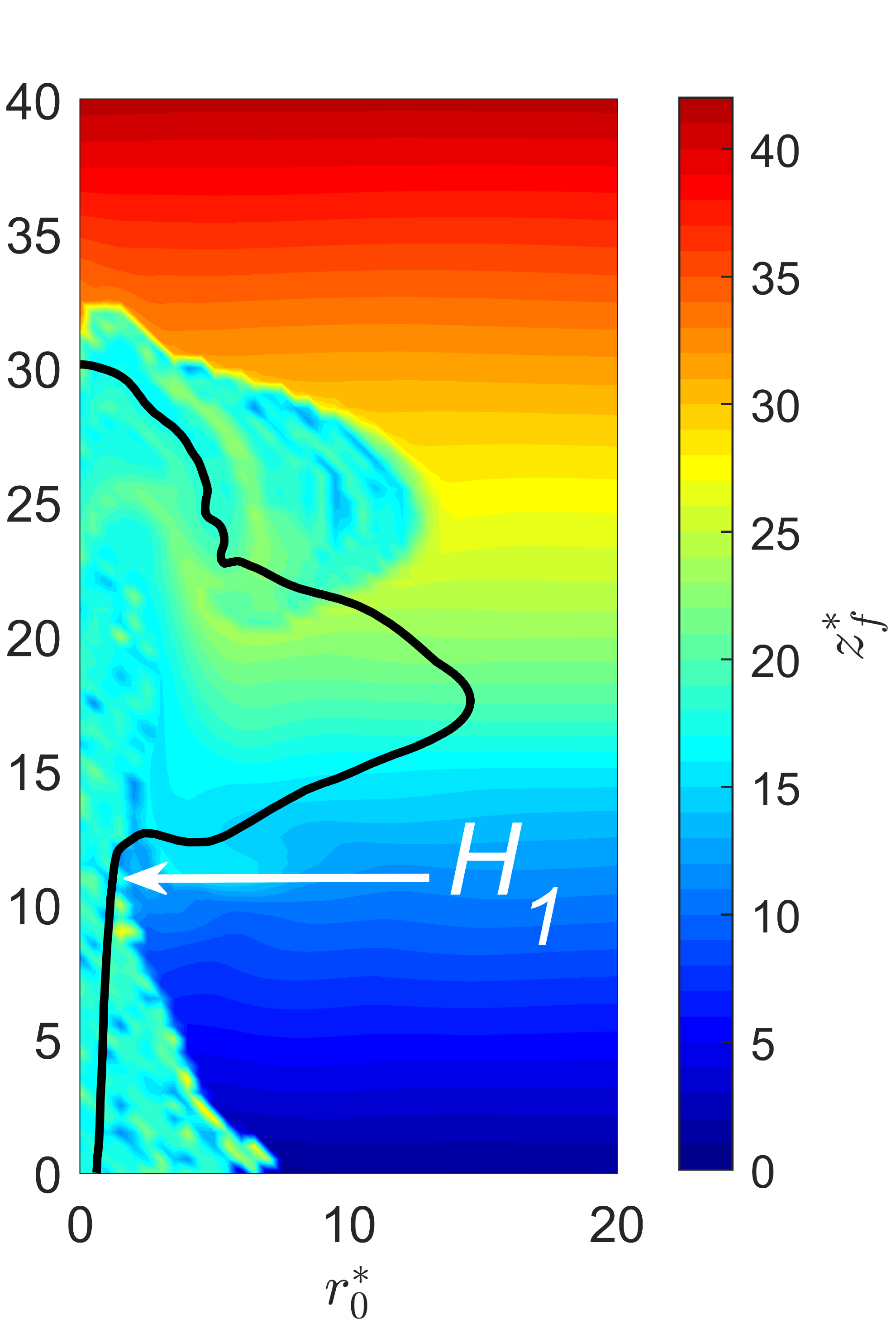}
    \caption{F-15 configuration.}
    \label{fig:zf_con_contornos_F15}
\end{subfigure}
\caption{Colour map of the final vertical dimensionless position ($z_f^*$) of the fluid particles at $t=100$~s for the G-15 (F-15) on the left (right), where the axis corresponds to the initial radial and vertical dimensionless coordinates of the particles ($r_0$ and $z_0$, respectively). We also show the fountain contour in magenta dashed line and indicate the limit height of region $H_1$ in both cases.}
\label{fig:zf_con_contornos}
\end{figure}

From the fluid particle trajectories, already calculated using the FTLE calculation procedure, we compute the volume of entrainment fluid at a given time horizon $t_f$ as follows. First, we compute each fluid particle's final position ($r_f$ and $z_f$ for the radial and final vertical position, respectively) at $t_f$. In this section, once again we consider $t_f=100$~s, as the flow is fully developed at this stage. In Fig.~\ref{fig:zf_con_contornos} we show the colour map of $z_f^*$ for the G-15 and F-15 configurations (the F-18 configuration is similar) as a function of the initial position $(r^*_0,z^*_0)$ (where dimensionless lengths are obtained dividing by $D$).

Given the azimuthal symmetry around the $z$ axis (due to the smoothing process), it is easy to compute the volume of ambient fluid that was within region $H_1$ at $t=0$ and arrived at $H_2$ (or a higher region) at $t_f$. We refer to this volume as $V_{rmv}(t)$, and to the volume of fluid that entered the domain through the inlet nozzle as $V_{in}(t)=q_{in}\cdot t$. In Tab.~\ref{tab:vol_removed} we show $V_{rmv}(tf)$ for three different ranges of initial vertical position, for the three configurations and the efficiency removal dimensionless parameter at $t_f$, $\xi_{rmv}(t_f)$, defined as $\xi_{rmv}(t_f)=V_{rmv}(t_f)/V_{in}(t_f)$.

From Tab.~\ref{tab:vol_removed} it is notable how turbulence (G-15 configuration) favours the entrainment of ambient fluid and then increases the efficiency in the removal of heavy ambient fluid. F-15 and F-18 performances are similar, revealing that heating the ejected fountain has no significant impact on the efficiency of the SIS, i.e., turbulence has a substantial effect on $\xi_{rmv}$, in contrast to the fountain temperature.

\begin{table}[hbt]
\centering
\caption{Volume of cold fluid removed from the environment at $t_f=100$~s, for the three configurations and for different strata height ranges, as a measure of the efficiency of the fountain to sink harmful fluid.}
\begin{tabular}{c|c|c|c}\hline
 & \multicolumn{3}{c}{$\xi_{rmv}$ ($\times10^{-4}$) at $t_f=100~\mathrm{s}$} \\
Configuration & $z^*\in (0,3)$ & $z^*\in (0,6)$ & $z^*\in (0,9)$ \\
\hline
G-15 & 3.44 & 5.71 & 9.27 \\
F-15 & 3.24 & 4.96 & 5.81 \\
F-18 & 3.13 & 4.85 & 5.89 \\
\hline
\end{tabular}
\label{tab:vol_removed}
\end{table}

In conclusion, let us suppose that the SIS device incorporates a heating system for the removed and ejected air. This makes the characteristic heights ($z_m$ and $z_{s}$) larger, as seen in both Fig.~\ref{fig:validation_heights} and Tab.~\ref{tab:alturas_regiones}. However, there is no growth in the removed volume of ambient fluid, as observed in Tab.~\ref{tab:vol_removed}, when comparing F-15 and F-18 performances. Furthermore, additional energy is required for the heating of the fluid, making the F-18 configuration even less efficient. Nonetheless, increasing the turbulence of the ejected fountain benefits the SIS efficiency, as shown in Tab.~\ref{tab:vol_removed} for the G-15 configuration. Although in this case, the characteristic heights are lower in comparison with the F-configurations (see Fig.~\ref{fig:validation_heights}), the spreading \textit{cloud} is located on safe heights for the crops ($H_2$ from Tab.~\ref{tab:alturas_regiones}), given the dimensions of the SIS (its diameter $D$) and the characteristic heights of the crops (e.g. fruit trees are bellow 6-8~$D$).

In summary, under the working conditions of our study, we conclude that: 1) increasing the turbulence benefits the efficiency of the SIS in the cold ambient fluid removal, and 2) temperature impact on the SIS performance is not relevant. However, we emphasise that further research on how to achieve the optimal conditions of the fountain is required for the optimal design of the SIS.

\section{\label{sec:conclusions} Conclusion}
In this work, we studied the flow structure of turbulent fountains in stratified media. For this purpose, we obtained the trajectories of particles initially located in different places within the environment domain. Since the fountain is turbulent, most of the trajectories are stochastic, and, consequently, there is significant uncertainty in the final positions. However, the average trajectories exhibit a clear structure with a primary downward flow stream exhibiting an oscillatory damped behaviour. A less significant stream promotes the particles to re-enter the up-down jet in agreement with some hypotheses made in previous models. Based on the FTLE analysis, we found the dominant LCSs that organize the flow in three different configurations. The main dynamic characteristics, such as the entrainment mechanisms and buoyancy force direction, were the elements to explain the most prominent structures. A more detailed study of the self-similarity features will be the scope of future work. 

Furthermore, based on the LCSs, we defined the surface delimiting the uprising fountain (referred to as the DAB), crossed by the fluid particle trajectories. From the DAB, we proposed a criterion for calculating the rate of entrainment (which takes place between the axial upwards stream and the environment) and re-entrainment (between the stream and the downflow) of fluid by the fountain. Moreover, as a novel contribution, we calculated the extended definition of the entrainment coefficient at advanced flow stages, where the fluid surrounding the uprising fountain is not quiescent. Due to inherent turbulent fluctuations and the requisite smoothing procedures, delimiting the DAB is not a trivial task. Hence, we averaged the azimuthal components, simplifying the analysis but keeping the most relevant characteristics. Finally, we applied the previous results to analyse the efficiency of the  SIS. We concluded that the turbulence of the ejected fountain is the critical feature that benefits its performance, referring to the harmful fluid that lies over the soil when radiation frost occurs, whereas the local temperature was found not to be a relevant feature. We estimate that the results obtained in the present work can be beneficial for constructing refined models focused on the description and prediction of the evolution of turbulent fountains.

\section*{\label{sec:ack} Acknowledgements}

The authors would like to thank PEDECIBA (MEC, UdelaR, Uruguay) and express their gratitude for the grant Fisica Nolineal (ID 722) Programa Grupos I+D CSIC 2018 (UdelaR, Uruguay). 

\section*{\label{sec:data} Data availability}

The data that support the findings of this study are available from the corresponding author upon reasonable request.

\section*{\label{sec:biblio} References}

\bibliography{biblio}

\end{document}